\documentclass[amsmath,amssymb,aps,prl,twocolumn,notitlepage,superscriptaddress,longbibliography,floatfix,10pt]{revtex4-2}

\usepackage{graphicx} 
\usepackage{amsmath}
\usepackage{amsbsy}
\usepackage{cancel}
\usepackage{enumerate}
\usepackage{mathtools}
\usepackage{amsfonts}
\usepackage{amssymb,dsfont,physics}
\usepackage{nicefrac}
\usepackage{xcolor}
\usepackage{comment}
\usepackage{scalerel}
\usepackage{nicefrac}
\usepackage[colorlinks=true,linkcolor=blue,citecolor=blue]{hyperref} 
\usepackage{tikz}
\usetikzlibrary{positioning}
\usetikzlibrary{automata,positioning}
\usetikzlibrary{fit,shapes.geometric}
\usetikzlibrary{decorations.pathmorphing}
\usetikzlibrary{arrows,matrix}
\usetikzlibrary{calc}
\usepackage{xcolor}
\usepackage[normalem]{ulem}

\usepackage{pict2e}

\definecolor{ccyan}{rgb}{0.55,0.95,0.8}
\definecolor{rred}{rgb}{1,0.2,0.2}

\DeclareRobustCommand{\lowerrighttriangle}{%
  \begingroup
  \setlength{\unitlength}{1ex}%
  \begin{picture}(1,1)
  \polyline(0,1)(0,0)(1,0)(0,1)(0,0.5)
  \end{picture}%
  \endgroup
}

\newcommand{\proj}[1]{\ket{#1}\!\bra{#1}}

\newcommand{\N}{\mathcal{N}}
\newcommand{\M}{\mathcal{M}}
\newcommand{\E}{\mathcal{E}}
\renewcommand{\P}{\mathcal{P}}
\newcommand{\F}{\mathcal{F}}
\newcommand{\D}{\mathcal{D}}
\newcommand{\U}{\mathcal{U}}
\newcommand{\C}{\mathcal{C}}
\newcommand{\R}{\mathcal{R}}
\newcommand{\Sw}{\mathcal{S}}
\newcommand{\T}{\mathcal{T}}
\newcommand{\id}{\mathbf{Id}}
\newcommand{\Id}{\mathds{1}}

\newcommand{\q}[1]{ q_{\scaleto{#1\mathstrut}{5pt}}}

\newcommand{\symsc}[2]{ #1_{\scaleto{#2\mathstrut}{5pt}}}
\newcommand{\scale}[2]{ #1_{\scaleto{#2\mathstrut}{5pt}}}
\newcommand{\su}[2]{ #1^{\scaleto{#2\mathstrut}{5pt}}}

\newcommand{\EE}{\tilde{\mathcal{E}}}
\newcommand{\FF}{\tilde{\mathcal{F}}}

\newcommand{\ee}{\tilde{E}}
\newcommand{\ff}{\tilde{F}}
\newcommand{\kk}{\tilde{K}}
\newcommand{\com}[2]{\left[#1, #2\right]}
\newcommand{\acom}[2]{\left\{#1, #2\right\}}

\newtheorem{result}{Result}

\newtheorem{definition}{Definition}

\newcommand{\sw}[1]{${\rm SWITCH}_{#1}$}

\begin{document}

\title{Probabilistic Channel Distillation via Indefinite Causal Order}
\author{Spiros Kechrimparis}
\email{skechrimparis@gmail.com}
\affiliation{School of Computational Sciences, Korea Institute for Advanced Study, Seoul 02455, South Korea}

\author{James Moran}
\email{jamesmoran@kias.re.kr}
\affiliation{Quantum Universe Center, Korea Institute for Advanced Study, Seoul 02455, South Korea}

\author{Hyukjoon Kwon}
\email{hjkwon@kias.re.kr}
\affiliation{School of Computational Sciences, Korea Institute for Advanced Study, Seoul 02455, South Korea}

\begin{abstract}
The quantum switch has been widely studied as a prototypical example of indefinite causal order in quantum information processing. However, the potential advantages of utilising more general forms of indefinite causal orders remain largely unexplored. We study higher-order switches, which involve concatenated applications of the quantum switch, and we demonstrate that they provide a strict advantage over the conventional quantum switch in the task of quantum channel distillation. Specifically, we show that higher-order quantum switches enable the probabilistic distillation of any qubit Pauli channel into the identity channel with nonzero probability. This capability contrasts with the conventional quantum switch, which allows only a limited set of Pauli channels to be distilled with nonzero probability. We observe that, counterintuitively, the distillation rate generally increases the noisier the channel is. 
We fully characterise the asymptotic distillation rates of higher-order superswitches for qubit Pauli channels.  
Finally, we prove a no-go result for multi-qubit generalisations. 
\end{abstract}

\maketitle

Emerging from the context of quantum gravity \cite{hardy_quantum_2007, hardy_quantum_2009}, it was observed that quantum theory is consistent with operations that are not applied in a definite causal order \cite{oreshkov_quantum_2012,chiribella_quantum_2013}, an observation that has sparked intense research since. The \emph{quantum switch} \cite{chiribella_quantum_2013}, the prototypical example of a supermap \cite{chiribella_transforming_2008,chiribella_quantum_2008} with indefinite causal order, has since found application in a variety of topics ranging from quantum computation \cite{araujo_computational_2014, taddei_computational_2021}, quantum metrology \cite{zhao_quantum_2020}, to quantum work extraction \cite{simonov_work_2022}.
Advantages for classical and quantum communication by using indefinite causal order have been shown in Refs.\@ \cite{ebler_enhanced_2018,salek_quantum_2018,delsanto_twoway_2018, caleffi_quantum_2020,procopio_communication_2019,procopio_sending_2020,sazim_classical_2021,chiribella_quantum_2021, mukhopadhyay_superposition_2020,das_quantum_2022,kechrimparis_enhancing_2024}. Since then, various experiments have been performed that demonstrate these findings; for a review, see Ref.\@ \cite{rozema_experimental_2024}.
In Ref.\@ \cite{chiribella_indefinite_2021} it was shown that noiseless communication is possible deterministically through an entanglement breaking channel which is unique, up to unitary equivalence, by inputting two copies of the channel in the quantum switch. 
Previously, probabilistic perfect transmission of quantum information was shown to be possible with the quantum switch for some Pauli channels \cite{salek_quantum_2018}.

Higher-order generalisations of the quantum switch have been introduced \cite{procopio_communication_2019,procopio_sending_2020,mukhopadhyay_superposition_2020,sazim_classical_2021,chiribella_quantum_2021,das_quantum_2022, kechrimparis_enhancing_2024}. The \emph{$N$-switch} superposes the ordering of $N$ channels through coherent control of all possible $N!$ orders \cite{procopio_communication_2019, procopio_sending_2020}. Superposition of the $N$ cyclic orders of $N$ depolarising channels channels has also shown advantages in communication \cite{chiribella_quantum_2021}. The \emph{switch of switches} has also been considered in \cite{das_quantum_2022,kechrimparis_enhancing_2024} and has demonstrated advantages in quantum communication and state discrimination. Although they have demonstrated advantages in some instances, in most cases these improvements are not always clear-cut. In addition, the coherence needed to implement higher-order switches scales badly, in general, further diminishing the returns and deeming their usefulness questionable. In this work, we first report on a task that certain types of higher-order switches can perform, while the quantum switch cannot, thus demonstrating that they can have a clear advantage over the quantum switch. The task at hand is \emph{exact probabilistic channel distillation}, which by analogy to other resource distillation scenarios \cite{bennett_purification_1996,bennett_concentrating_1996, kent_entangled_1998, horodecki_general_1999, lo_concentrating_2001,jane_purification_2002, horodecki_fidelity_2006,regula_oneshot_2019,streltsov_colloquium_2017,wu_quantum_2020,liu_optimal_2021,forster_distilling_2009,eftaxias_advantages_2023,naik_distilling_2023}, consists of turning multiple copies of a noisy channel into the identity or a unitary channel, with some nonzero probability.

We first notice that the quantum switch allows probabilistic exact channel distillation essentially for a set of measure zero in the set of unital channels, while for most channels it fails. We then show that by using higher-order switches, and specifically switches of switches, probabilistic distillation is possible with \emph{any} unital channel in dimension two. 
Counterintuitively, we find that the noisier a channel is, the better it performs in this distillation protocol.

As higher-order switches can be implemented with current technology \cite{chiribella_quantum_2019, rozema_experimental_2024}, our results can thus be employed to achieve noiseless classical and quantum communication in $d=2$ with any channel. From a foundational perspective, and in view of the Choi–Jamiołkowski isomorphism \cite{choi_completely_1975,jamiolkowski_linear_1972}, our results have implications for entanglement sharing between two parties, where the first party prepares an entangled state, keeps one part of the system, and sends the second part to the other party over a quantum channel. This task is impossible over an entanglement-breaking channel, as it maps any state to a separable state.  
With access to higher-order switches, any unital channel in dimension $d=2$ can be used to distribute entanglement with nonzero probability, including all channels from the set of entanglement-breaking channels, enabling applications that would otherwise be impossible.

\textit{Quantum switch and superswitches.---}  
The quantum switch \cite{chiribella_quantum_2013} is a supermap that superposes the ordering between the actions of two channels $\E$ and $\F$, conditioned to a $2$-dimensional switch state $\omega$. The resulting channel is defined as
\begin{equation}
S_\omega(\E,\F) = \sum_{i,j}S_{ij} (\rho\otimes \omega)S_{ij}^\dag,
\end{equation} 
with $S_{ij} = E_i F_j \otimes \proj{0}_C+F_j E_i \otimes \proj{1}_C $. Here, $E_i, F_j$ denote the Kraus operators of the channels $\E$ and $\F$, respectively, and $\ket{0}\bra{0}_C$ and $\ket{1}\bra{1}_C$ denote the projectors acting on the ancilla state. $S_{ij}$ are Kraus operators satisfying $\sum_{ij} S_{ij}^\dagger S_{ij} = \Id$ when both ${\cal E}$ and ${\cal F}$ are Pauli channels, represented as ${\cal P}_{\vec p}(\rho) = p_0 \rho + p_1 X \rho X + p_2 Y \rho Y + p_3 Z \rho Z$ with $\vec p = (p_1, p_2, p_3)$ and $p_0 =1-p_1-p_2-p_3$ \cite{chiribella_quantum_2013,chiribella_indefinite_2021}. In particular, by taking the ancilla state to be $\omega = \ket{+}\bra{+}$ with $\ket{\pm} = (\ket{0} \pm \ket{1})/\sqrt{2}$, we obtain
\begin{align}
S_\omega(\E, \F) &= {\cal C}_+(\E, \F) (\rho) \otimes \ket{+}\bra{+}  \notag \\
&+{\cal C}_-(\E, \F)(\rho) \otimes \ket{-}\bra{-}, \label{eq: switch operation}
\end{align}
where 
$$
\begin{aligned}
    {\cal C}_{\pm}(\E, \F)(\rho) &= \sum_{i,j} [E_i, F_j]_\pm \rho [E_i, F_j]_\pm^\dagger,
\end{aligned}
$$
with $[A, B]_\pm = AB \pm BA$ denoting the anticommutator and commutator. $\C_{\pm}(\E,\F)(\rho)$ can be expressed as probabilities $q^{(1)}_\pm$, multiplied by channels $\E^{(1)}_{\pm}(\rho)=\C_{\pm}(\E,\F)(\rho)/q^{(1)}_\pm$, where $q^{(1)}_\pm=\tr{\C_{\pm}(\E,\F)(\rho)}$. Consequently, a measurement of the ancilla state in the basis $\{ \ket{+}, \ket{-} \}$, results in the separation of the two channels. The case with $\E=\F$ is most commonly studied, which we also assume in this work. 

\begin{figure}[!t]
	\includegraphics[width=0.8\linewidth]{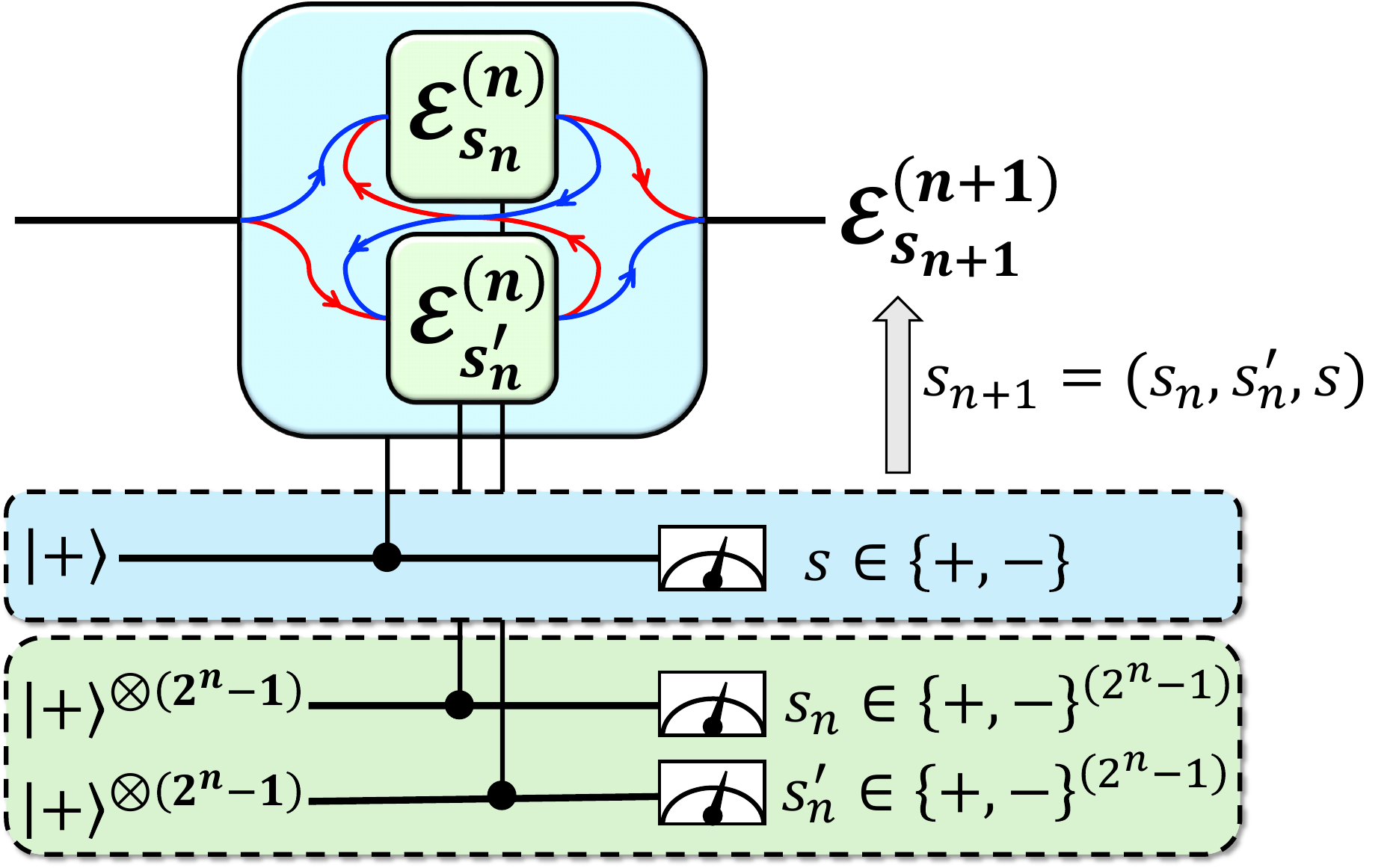}
	\caption{The $(n+1)$-order superswitch is generated by supersposing the ordering two $n$-order superswitches. The ancilla qubits (in green) control the ordering of channels inside the $n$-order superswitches, while another ancilla (blue) controls their ordering. \label{fig:protocol}}
\end{figure}

Higher-order generalisations of the quantum switch have also been defined \cite{procopio_communication_2019,procopio_sending_2020,mukhopadhyay_superposition_2020,sazim_classical_2021,chiribella_quantum_2021,das_quantum_2022}. A particular case are switches of switches, also known as superswitches. An instance with synchronised ordering on the inside switches was studied in Ref.\@ \cite{das_quantum_2022}, while the general case was introduced and studied in Ref.\@ \cite{kechrimparis_enhancing_2024} and applied to the problem of quantum state discrimination. The $n$-order superswitch effectively evaluates all possible nested expressions of commutators and anticommutators of the Kraus operators of the channels.
We denote superswitches by \sw{n}, with $n$ denoting their order. \sw{n} is obtained by applying the switch operation in Eq.\@ \eqref{eq: switch operation} to the two lower-order superswitches \sw{n-1} (see Fig.\@ \ref{fig:protocol}). The case $n=0$ corresponds to the application of no superswitch, i.e. just the channel, and $n=1$ is the case of the quantum switch. To implement \sw{n}, $2^n$ copies of the channel $\E$ and $2^n-1$ ancilla qubits are needed. The role of the ancillas is to control the ordering of all nested switches.

At each order, the action of \sw{n} can be effectively described by the tuples $\Omega^{(n)}_{s_n}=\{q^{(n)}_{s_n}, \E^{(n)}_{s_n} \}$, with $\E^{(n)}_{s_n}$ denoting channels and $q^{(n)}_{s_n}$ their probability of occurrence; $s_n$ denotes a string of length $2^{n}-1$ consisting of `$\pm$' outcomes corresponding to the measurements on the ancilla qubits.
A superswitch of any order $n$ can be evaluated from the previous order $n-1$ (see Fig.\@ \ref{fig:protocol}). Explicitly, we find for the tuples $\Omega^{(n)}_{s_n}=\{q^{(n)}_{s_n}, \E^{(n)}_{s_n} \}$,
\begin{align}
     \E_{s_n}^{(n)} &= \C_{s}\left(\E^{(n-1)}_{s_{n-1}} , \E^{(n-1)}_{s_{n-1}'} \right)/ q^{(n)}_{s_n} \,,\notag \\
     q^{(n)}_{s_n} &= \tr\left[\C_{s}\left(\E^{(n-1)}_{s_{n-1}} , \E^{(n-1)}_{s_{n-1}'} \right)(\rho) \right] \,,\label{eq: iteration superswitches}
\end{align}
where $s_n = s_{n-1}  s_{n-1}^\prime  s$ denotes the concatenation of strings $s_{n-1}, s_{n-1}^\prime$, and $s$. 
At order $n=0$ we only have the channel, and thus the initial conditions are $q^{(0)}=1$ and $\E^{(0)}=\E$. Iterating once, we get the two channels of the conventional quantum switch, while a second iteration gives the first higher-order generalisation $\Omega^{(2)}_{\mu \nu s}=\{q^{(2)}_{\mu \nu s}, \E^{(2)}_{\mu \nu s} \}$ consisting of eight channels and their respective probabilities of occurrence, and where $\mu,\nu$ are strings of `$\pm$' of length one.
In Appendix \ref{app: n-order superswitches} we list all the expressions $\Omega^{(2)}_{\mu \nu s}$ as well as the general expression for the $n$-order superswitch.

\textit{Quantum channel distillation via higher-order quantum switches.---}
Resource distillation is the task of converting $k$ less-resourceful copies of a certain resource into $l$ copies with higher resource content, where $k\geq l$. Examples include protocols to distil entanglement, coherence, and nonlocality \cite{bennett_purification_1996,bennett_concentrating_1996, kent_entangled_1998, horodecki_general_1999, lo_concentrating_2001,jane_purification_2002, horodecki_fidelity_2006,regula_oneshot_2019,streltsov_colloquium_2017,wu_quantum_2020,liu_optimal_2021,forster_distilling_2009,eftaxias_advantages_2023,naik_distilling_2023}. Such a task can be exact or approximate depending on whether the target resource is achieved exactly or up to a certain approximation. In addition, such protocols can be deterministic or probabilistic, depending on whether the conversion is successful with certainty or only with some nonzero probability.
In this work, we consider the task of exact distillation of multiple copies of a given channel $\E$ into the identity or a unitary channel $\U$. In the latter case, applying the inverse unitary recovers the identity channel as well. Thus, we will consider exact probabilistic channel distillation protocols. 
Our proposed channel distillation protocols employ \sw{n}, which require $2^n$ copies of the channel $\E$ in order to implement. 
\begin{definition}
Let $D^{(n)}$ be the collection of strings of $2^{n}-1$ plus or minus outcomes such that the corresponding channel is the identity or a unitary channel. That is, with $\tilde{s}\in D^{(n)}$ we either have $\E_{\tilde{s}}^{(n)} =\id$ or $\E_{\tilde{s}}^{(n)} = \U $, where $\U(\rho)=U\rho U^\dagger$ denotes an arbitrary unitary channel. Then, the \emph{distillation rate} for a channel $\E$, $\R^{(n)}_{\E} \in [0,1]$, of a protocol based on \sw{n} is given by
    \begin{align}
        \R^{(n)}_{\E} = \sum_{\tilde{s}\in D^{(n)}} q^{(n)}_{\tilde{s}} \,.
    \end{align}
\end{definition}

In Ref.\@ \cite{chiribella_indefinite_2021} it was shown that the channel $\E_{XY}=\frac{1}{2}(X\rho X+Y \rho Y)$ is the unique channel that is both entanglement-breaking and therefore useless for quantum communication, but leads to \emph{perfect} communication with unit probability when combined with itself in the quantum switch.
In our setting, this translates to the statement that there is a unique entanglement-breaking channel, up to unitary transformations, that can be exactly purified deterministically with the quantum switch. More generally, any channel with zero portion of the identity, $p_0=0$, can be probabilistically purified with the quantum switch. 

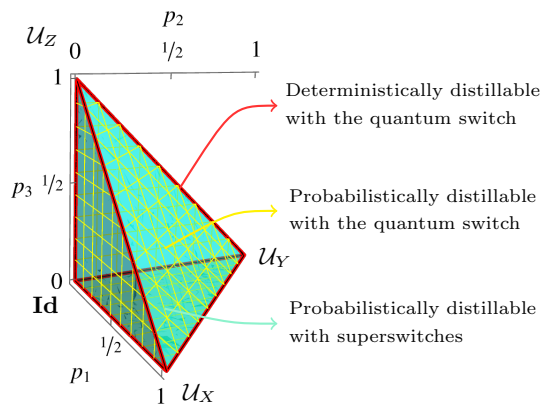
\begin{figure}[!t]
	\begin{tikzpicture}[node distance=1cm, thick,square/.style={regular polygon,regular polygon sides=4,minimum size=0cm}, main/.style={draw, circle,minimum size=0.5cm}]
		\node[] (1) {\includegraphics[trim={8mm 0 8mm 0}, width=0.5\columnwidth]{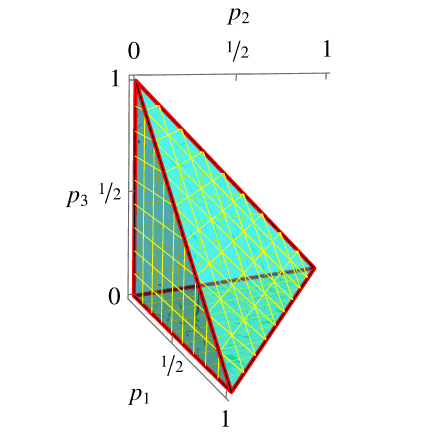}}; 
		\node[] [below left = -1.85cm and -1.1cm of 1] (2) {$\id$};
		\node[]  [below left = -0.7cm and -3.2cm of 1] (3) {$\U_X$}; 
		\node[]  [below left = -2.5cm and -4.2cm of 1] (4) {$\U_Y$}; 
		\node[] [above left = -0.8cm and -1.1cm of 1] (5) {$\U_Z$};
  
		\draw [->, rred] (0.3,0.3) .. controls (1,1.4) .. (1.6,1.4);
    	\draw [->, yellow] (0.1,-0.5) .. controls (1,0) .. (1.6,0);   \draw [->, ccyan] (0.1,-1.3) .. controls (1,-1.5) .. (1.6,-1.5);
     
        \node[align=left] at (3.4,1.4) {\scriptsize Deterministically distillable\\\scriptsize with the quantum switch};
    	\node[align=left] at (3.4,0) {\scriptsize Probabilistically distillable\\\scriptsize with the quantum switch};
    	\node[align=left] at (3.4,-1.5) {\scriptsize Probabilistically distillable\\ \scriptsize with superswitches};
	\end{tikzpicture} 
	
	\caption{The tetrahedron of the set of Pauli channels (see Appendix \ref{app: geometric picture of Pauli channels in d=2}). On the edges are channels that can be distilled with the quantum switch deterministically (red lines), while channels on the faces can be distilled probabilistically (yellow meshed surfaces). In the interior are channels that can only be distilled with superswitches (cyan). $\U_P$ denotes a unitary channel $P\ldots P^\dagger$ with $P = X,Y,Z$.  \label{fig: quantum switch channels}}
\end{figure}

By considering probabilistic protocols, more channels can be purified. As the set of Pauli channels ${\cal P}_{\vec p}(\rho) = p_0 \rho + p_1 X \rho X + p_2 Y \rho Y + p_3 Z \rho Z$ with $\vec p = (p_1, p_2, p_3)$ forms a tetrahedron in $(p_1,p_2,p_3)$ space (see Appendix \ref{app: geometric picture of Pauli channels in d=2}), the set of channels that can be distilled by the quantum switch lie at the boundary of the tetrahedron, that is, the four faces and the channels at the edges. Inside the tetrahedron, no channel can be distilled with the quantum switch (see Fig.\@ \ref{fig: quantum switch channels}). We note that even though the three axes of the tetrahedron consist of channels that cannot be distilled by direct application of the quantum switch (or even the protocols we will propose), a simple application of one of the Pauli matrices $\sigma_i$ with the same index $i$ as one of the zero $p_i$'s in the definition of the channel, before inputting in the quantum switch, maps it to a channel with nonzero $p_i$'s. This straightforward pre-processing is a technique that has also been employed in \cite{mitra_improvement_2023,wu_general_2024}. 

\vspace{0.5mm}
\textit{Distillation Protocol with \sw{2}.---} Our first result shows that \sw{2} allows purification of all channels inside the tetrahedron.
\begin{result}
	 Any qubit Pauli channel inside the tetrahedron can be exactly distilled to the identity with a nonzero probability when four copies of the channel are combined in the second-order superswitch. Moreover, a Pauli channel $\E=\P_{\vec{p}}$ with $p_i\neq 0\, \forall i$, has distillation probability:
     \begin{align}
         \R^{(2)}_\E=\q{--+}^{(2)}= 4 \big(p_1^2 p_2^2+p_2^2 p_3^2+p_3^2 p_1^2\big)\,.
     \end{align} 
 \end{result}

The proof is given in Appendix \ref{app: proof of results 1 and 2}. We note that this probability is nonzero, i.e. $\q{--+}>0$, for all channels except for the trivial case of $\E_{\vec{p}}=\id$ and channels where two of the $p_i$, with $i\in \{1,2,3\}$, are equal to zero. In the latter case, applying one of the Pauli matrices $\sigma_i$ with the same index $i$ as the zero $p_i$ to the channel before inputting in the superswitch maps the channel to a channel with only one zero $p_i$, thus achieving a nonzero probability $\q{--+}$ according to Result 1.

Result 1 is a consequence of the fact that the channel corresponding to the outcome `$--+$' for \sw{2} after measurement in the ancilla qubits is the identity channel, i.e. $\symsc{\E}{--+}^{(2)}=\id$ (see Appendix \ref{app: proof of results 1 and 2}). The reason that this is possible for \sw{2} and not the quantum switch is due to the following two facts: (i) for any Pauli channel input in the quantum switch, a `$-$' result on the ancilla results in a Pauli channel with $p_0=0$, and (ii) for any Pauli channel with $p_0=0$ input in the quantum switch, a `$+$' outcome leads to the identity channel. Thus, even though the quantum switch can lead to the identity channel for the input channels with $p_0=0$, or any of the other three faces of the tetrahedron by pre-multiplying with a unitary, it cannot lead to the identity for any channel in the interior: the quantum switch cannot distil any channel inside the tetrahedron. In contrast, as \sw{2} is a switch of switches, both (i) and (ii) can simultaneously occur, leading to nonzero distillation in the interior.

Returning to the distillation rate in Result 1, we make the counterintuitive observation:  
\begin{result}
    Given two Pauli channels $\E=\P_{\vec{p}}$ and $\F=\P_{\vec{q}}$ such that $p_i\geq q_i\,, \, \, \forall i\in\{1,2,3\}$, we have $\R^{(2)}_\E\geq\R^{(2)}_\F$, with strict inequality if for at least one value of $i$, we have $p_i > q_i$. That is, the noisier channel $\E$ will have a higher distillation rate with the \sw{2} protocol.
\end{result}

\textit{Distillation Protocol with \sw{n}.---}
For distillation protocols with \sw{n}, a similar argument holds. As these are nested switches, any sequence of outcomes $s$ that ends in `$\ldots--+$' implies that the channel obtained is the identity. However, in practice, there will be more contributions to the distillation rate. 
To account for all contributions in protocols with \sw{n}, we can set up recurrence relations that can be iterated to obtain the distillation rates. We first describe the construction in the interior before examining the faces of the tetrahedron. At order $n$ we can write the channel of \sw{n} as 
\begin{align}
    \Sw^{(n)} =& \alpha^{(n)} \id \otimes \rho_{\id}+\beta^{(n)} \N_{\vec{q}^{\,(n)}}  \otimes   \rho_{\N}  \notag \\
    &+\gamma^{(n)} \P_{\vec{p}^{\,(n)}}  \otimes   \rho_{\P} \,. \label{eq: effective superswitch maintext}
\end{align}
where $\rho_{\id}, \rho_{\N}, \rho_{\P}$ denote unnormalised states with orthogonal supports that gather the outcomes $o_{\id}, o_{\N}, o_{\P}$ of measurements on the ancillas that lead to the identity channel $\id$, a channel with zero portion of the identity $\N$, i.e. $q^{(n)}_{0}=0$, and a general Pauli channel with $p^{(n)}_{0} \neq 0,1$, respectively. For instance, in the case of the second-order superswitch, we have $\rho_{\id} = \ketbra{\scriptstyle --+},  
    \rho_{\N}=\ketbra{\scriptstyle ++-}+\ketbra{\scriptstyle-+-}+\ketbra{\scriptstyle+--}+\ketbra{\scriptstyle---}, \rho_{\P}=\Id -\rho_{\id}-\rho_{\N}$ (see Appendix \ref{app: n-order superswitches}).
The fact that it is possible to summarise all different channels into essentially three channels, follows from the fact that for Pauli channels $\P_{\vec{p}_i}$ we have
$\sum_i \lambda_i \P_{\vec{p}_i} = \left(\sum_i \lambda_i\right) \P_{\vec{p}^\prime}$, with $\vec{p}^{\prime}=\nicefrac{\sum_i \lambda_i \vec{p}_i}{\sum_i \lambda_i}$.

Given that superswitches are switches of switches, and that the action of the superswitch at order $n$ is assumed to take the form in Eq.\@ \eqref{eq: effective superswitch maintext}, we can explicitly evaluate the superswitch at order $n+1$. In practice, this means that for each combination of terms we need to evaluate the contribution from the anticommutator and commutator from Eq.\@ \eqref{eq: iteration superswitches}. There are nine possible combinations, each of which leads to two possible outcomes so in principle we will have 18 terms at order $n+1$. However, by noticing that combining the identity with any channel leads only to one outcome, i.e. the channel itself, and that the ordering of combining two channels does not matter, we find nine unique terms, which can be rearranged and written in the form $\Sw^{(n+1)} = \alpha^{(n+1)} \id \otimes \rho_{\id}+\beta^{(n+1)} \N_{\vec{q}^{\,(n+1)}}  \otimes   \rho_{\N} +\gamma^{(n+1)} \P_{\vec{p}^{\,(n+1)}}  \otimes   \rho_{\P} $. The equations for $\alpha^{(n+1)}, \beta^{(n+1)}, \gamma^{(n+1)}, \vec{q}^{\,(n+1)},  \vec{p}^{\,(n+1)}$ define recurrence relations that can be iterated to obtain any order superswitch (see Appendix \ref{app: Recurrence Relations and Asymptotic Distillation Rate}). Even though we cannot solve the recurrence relations exactly, we look for their fixed points. We find exactly one solution: $ \alpha^{(\infty)} = \nicefrac{(2-\sqrt{3})}{4}, \beta^{(\infty)} = \nicefrac{\sqrt{3}}{4} , \gamma^{(\infty)}=\nicefrac{1}{2}$, $q^{(\infty)}_{1}=q^{(\infty)}_{2}=q^{(\infty)}_{3} = \nicefrac{1}{3}$ and $p^{(\infty)}_{1}=p^{(\infty)}_{2}=p^{(\infty)}_{3} = \nicefrac{(3-\sqrt{3})}{6}$. Thus, the asymptotic distillation rate is $\R^{(\infty)}_{\text{in}}= \nicefrac{(2-\sqrt{3})}{4} \approx 0.067$. Note that the asymptotic distillation rate does not always imply the highest achievable distillation rate (see Appendix \ref{app: Asymptotic distillation rates vs
finite-order distillation rates}).

The case of channels on the faces of the tetrahedron can be examined similarly. In this case, unitary channels can also contribute to the distillation rate, which changes the form of the recurrence relation. We find that the asymptotic distillation rate on the base of the tetrahedron, $\R^{(\infty)}_{\text{base}}$, is different to the rate on the other three faces, $\R^{(\infty)}_{\,\lowerrighttriangle{}}$ (see Appendix \ref{app: Recurrence Relations and Asymptotic Distillation Rate}).
\begin{figure}[!t]
	\includegraphics[width=1\columnwidth,trim={0.5cm 0.7cm 0 0},clip]{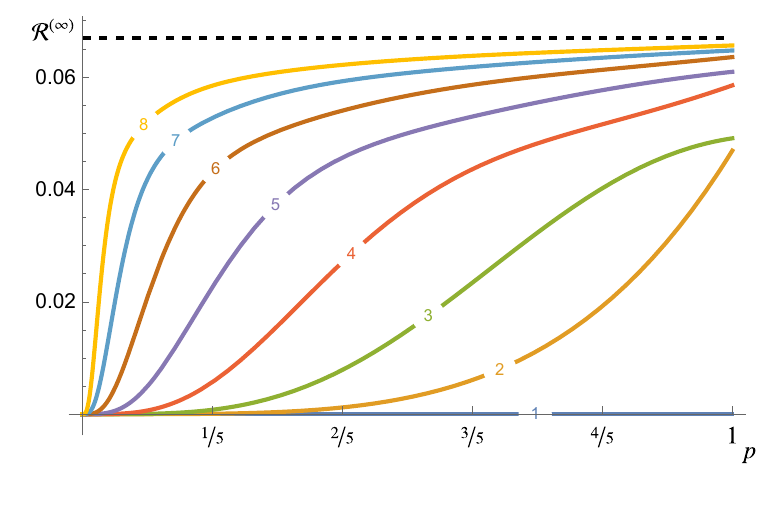}
	\caption{The distillation rates $\R^{(n)}$ achieved by the superswitches from orders one to eight. The quantum switch can not distill for any value of $p$, while each superswitch achieves a higher distillation rate for all values of $p\in[0,1]$, in comparison to all previous-order superswitches.} 
    \label{fig: communication rates}
\end{figure}
In summary, for the full set of Pauli channels, we have:
\begin{result}
    The asymptotic distillation rate inside the tetrahedron is $\R^{(\infty)}_{\text{in}}=\nicefrac{(2-\sqrt{3})}{4}\approx 0.067$. On the three faces of channels with nonzero identity part, i.e. $p_0\neq0$, the distillation rate is $\R^{(\infty)}_{\,\lowerrighttriangle{}}=\nicefrac{(3-\sqrt{3})}{4}\approx 0.396$, while on the face of channels with $p_0=0$, it is $\R^{(\infty)}_{\text{base}}=\nicefrac{1}{4}$. 
\end{result}

\textit{Examples.---} We now demonstrate our results for the depolarisation channel, $\D(\rho) = \left(1-\frac{3p}{4}\right)\rho + \frac{p}{4}\left(X\rho X + Y\rho Y + Z\rho Z \right)$ with $p\in[0,\nicefrac{4}{3}]$.  Distillation is impossible through a depolarisation channel with $p\neq \nicefrac{4}{3}$ with the quantum switch.
However, given access to \sw{2}, distillation becomes possible for all values of $p$. Specifically, the distillation rate is $\scale{\R}{\text{dep}}^{(2)}(p)=\nicefrac{3p^4}{64}$, which is nonzero for all values of $p$ and, moreover, it is a monotonically increasing function of the channel noise parameter $p$, demonstrating results 1 and 2. By iterating the recurrence relations (see Appendix \ref{app: distillation rate for n-order superswitches}), the distillation rates of higher-order superswitches further increase the distillation rate for all values of $p\in[0,1]$. In Fig.\@ \ref{fig: communication rates}, we show a plot of the distillation rates for \sw{n} with $n\leq 8$.
 The depolarisation channel is entanglement-breaking for all values of the noise parameter $p \in [\nicefrac{2}{3}, \nicefrac{4}{3}]$ (see Appendix \ref{app: geometric picture of Pauli channels in d=2}). Interestingly,  this is the range where the distillation rate is the highest for all orders $n$, echoing Result 2.

\textit{No-go result for multi-qubit Pauli channels.---} So far, we have discussed qubit channels. It is worth asking whether the above results can be generalised to higher dimensions. In Ref.~\cite{kechrimparis_enhancing_2024} superswitches were also defined in any dimension $2^l$ with $l\geq1$ and dimension $d=4$ was discussed in detail. We have the following negative result.
\begin{result}
    In any dimension $d=2^l$ with $l>1$, no superswitch of any order can lead to noiseless communication for all Pauli channels.
\end{result}
 Thus, Results 1 and 2 no longer hold for higher dimensions. The proof is given in Appendix \ref{app: Results 1 and 2 do not hold in d>2}, however, we offer some intuition as to why this is the case in general. We recall that the proof of Result 1 was based on two facts: (i) for any channel, a minus result on the ancillas gives a channel with $p_0=0$, and (ii) for any Pauli channel with $p_0=0$, a plus outcome leads to the identity channel. While (i) still holds, (ii) fails. 
 This follows from the commutators and anticommutators of tensor products of Pauli matrices.

\textit{Discussion.---} We have shown that exact channel distillation is possible probabilistically for any channel in dimension $d=2$, including the set of entanglement breaking channels. This proves that all channels can be used for perfect classical and quantum communication if two parties can implement at least the second-order superswitch, and have access to a sufficient number of copies of the communicated states and channel. 
Owing to the Choi–Jamiołkowski isomorphism \cite{choi_completely_1975,jamiolkowski_linear_1972}, access to higher-order superswitches thus implies that any channel in dimension $d=2$ can be used to distribute entanglement with nonzero probability, including all channels from the set of entanglement-breaking channels (see Appendix \ref{app: The Choi-Jamiołkowski isomorphism and entanglement sharing}). 
Useless channels from the perspective of sharing entanglement, even those that remain useless after using the quantum switch, can thus be used to distribute entanglement given access to higher-order superswitches \cite{mitra_improvement_2023}.
We have also found that, counterintuitively, the closer a channel is to the identity channel, the worse it performs at this task. 

Even though we have focused our discussion on Pauli channels, the case of non-unital channels $\N=\sum_{i} K_i \rho K_{i}^\dagger$ can be covered by employing twirling \cite{nielsen_simple_2002,gross_evenly_2007}. Each copy of the channel becomes a depolarisation channel with parameter $\eta=(\sum_i \abs{\tr K_i ^2}-1)/3$, leading to the distillation rate $\R_{\N}^{(2)} = 3 \eta^4/4$.  While this comes at a great resource cost (see Appendix \ref{app: Twirling channels and application to superswitches}), channel-specific-twirling protocols may be able to reduce that overhead. 

There are a number of interesting avenues for further research. 
Owing to the fact that the quantum switch has been implemented in practice in a number of different experiments, a direct adaptation of the setups to implement the second-order superswitch appears within reach, thus making our proposal for perfect quantum communication directly implementable with current technology.

Another interesting direction is to focus on dimensions $d>2$ with the aim of designing switch-like supermaps that can achieve noise-free communication for all unital channels. The algebraic structures related to the superswitches give some intuition on how to approach this problem, but whether such generalisations lead to physically allowable maps remains open.

Finally, our results assume that a perfect implementation of the superswitches is possible, which, however, may not be realistic in practical implementations. A promising future direction is to revisit our results in the case where the ancilla qubits are assumed to interact with their environment, thus not being in perfect coherent superpositions.
Quantifying the trade-off between coherence and fidelity under realistic noise models, would enable analysis of our protocol in practical settings.

\begin{acknowledgments}
	\textit{Acknowledgements. ---} S.K., J.M. and H.K. are supported by KIAS individual grant numbers CG086202 (S.K.), QP088702 (J.M.), and CG085302 (H.K.) at the Korea Institute for Advanced Study.
\end{acknowledgments}

\bibliographystyle{apsrev4-1}
\bibliography{switch_dist_bib.bib}

\appendix

\section{$n$-order superswitches}
\label{app: n-order superswitches}
The $n$-order superswitch can be iteratively defined as a switch of two $(n-1)$-order switches \cite{kechrimparis_enhancing_2024}. By denoting the Kraus operators of the $n$-order superswitch as $K^{(n)}$, these are explicitly defined in terms of those of the previous orders as
\begin{align}
	K^{(n)}_{\mathbf{i}_{n-1} \mathbf{i}^\prime_{n-1}}&=K^{(n-1)}_{\mathbf{i}_{n-1}}  \kk^{(n-1)}_{\mathbf{i}^\prime_{n-1}} \otimes \ketbra{0}_{2^{n}-1} \notag \\
	&\hspace{10mm}+ \kk^{(n-1)}_{\mathbf{i}^\prime_{n-1}} K^{(0)}_{\mathbf{i}_{n-1}} \otimes \ketbra{1}_{2^{n}-1} \notag \\
	&= \frac{1}{2} \left\{K^{(n-1)}_{\mathbf{i}_{n-1}}, \kk^{(n-1)}_{\mathbf{i}^\prime_{n-1}} \right\}\otimes\Id \notag \\
	&\hspace{10mm}+ \frac{1}{2}\left[K^{(n-1)}_{\mathbf{i}_{n-1}}, \kk^{(n-1)}_{\mathbf{i}^\prime_{n-1}}\right]\otimes Z,
\end{align}
where the index of the control qubit denotes that at order $n$ there are $2^{n}-1$ control qubits in total. The index symbol in bold, $\mathbf{i}_{n-1}$, is shorthand for the list of indices $\mathbf{i}_{n-1}\equiv i_1 i_2 \cdots i_{2^{n-1}}$ associated with the Kraus operators of the superswitch of order $n-1$. We use the convention $K^{(0)}=E_i, \tilde{K}^{(0)}=F_j\, \ldots$, that is, at order zero we only have the Kraus operators of the channels to be combined in a superswitch.
By expanding the nested expressions that consist of all commutators and anticommutators of all pairs of commutators and anticommutators generated by the Kraus operators of the channels, we get the following expression for \sw{n}
\begin{align}
	&\Sw^{(n)}_{\omega_1\cdots \omega_{2^{n}-1}}(\E,\F\cdots)= \frac{1}{2^{2^{n+1}-2}}\sum_{\substack{i_1,\cdots,i_{2^{n}-1}=\pm\\
    i,j,\ldots=1,\ldots,4}} \notag \\
	& \,\,\Bigg(\left[\cdots,\left[\left[E_i,F_j\right]_{i_1},\left[\EE_k,\FF_l\right]_{i_2}\right]_{i_3}\cdots\right]_{i_{2^{n}-1}} \rho \notag \\
	&\hspace{20mm}\times\left[\cdots,\left[\left[E_i,F_j\right]_{i_1},\left[\EE_k,\FF_l\right]_{i_2}\right]_{i_3}\cdots\right]_{i_{2^{n}-1}} \notag \\
	&\hspace{15mm}\otimes P_{i_1} \omega_1 P_{i_1} \otimes \cdots \otimes P_{i_{2^{n}-1}} \omega_{2^{n}-1} P_{i_{2^{n}-1}} \Bigg) \,, \label{eq: general switch}
\end{align}
with $P_+=\Id$ and $P_-=Z$. Explicit expressions for a given order $n$ can be evaluated iteratively \cite{kechrimparis_enhancing_2024}.

 In this framework the first-order superswitch is the quantum switch, which given two channels $\E$ and $\F$ with Kraus operators $E_i$ and $F_j$, respectively, is
\begin{align}
&\Sw^{(1)}_{\omega}(\E,\F)(\rho) = \sum_{i,j} K^{(1)}_{ij} \rho \otimes \omega K^{(1)}_{ij}  \notag \\	
	&\hspace{10mm}=\frac{1}{4} \sum_{i,j} \left\{E_i, F_j \right\}\rho\left\{E_i, F_j \right\}^\dagger\otimes \omega \notag \\
	&\hspace{12mm} +  \frac{1}{4} \sum_{i,j} \left[E_i, F_j\right]\rho \left[E_i, F_j\right] ^\dagger \otimes Z \omega Z \,, \label{eq: zero order ss- coms and anticoms}
\end{align}
matching the expression in the main text, Eq.\@ \eqref{eq: switch operation}.
The last expression shows that the quantum switch effectively evaluates channels that follow from commutator and anti-commutators of the Kraus operators of the channels. Non-trivial effects occur whenever some of the Kraus operators do not commute with each other.

We now explicitly derive the second-order superswitch, which is crucial for the results in the main text. Specifically, with four channels $\E$, $\F$, $\EE$ and $\FF$ that have Kraus operators $E_i$, $F_j$, $\tilde{E_k}$ and $\tilde{F_l}$ respectively, we initially define two ordinary quantum switches, that is, two first-order superswitches. Their Kraus operators, $K^{(1)}_{ij}$ and $\kk^{(1)}_{ij}$, are explicitly
\begin{align}
	K^{(1)}_{ij}&=E_i F_j \otimes \ketbra{0}_1+ F_j E_i \otimes \ketbra{1}_1 \notag \\
	\kk^{(1)}_{ij}&=\ee_i \ff_j \otimes \ketbra{0}_2+ \ff_j \ee_i \otimes \ketbra{1}_2 \,,
\end{align}
with the subscripts differentiating between the different ancilla qubits that control the ordering of the channels $\E, \F$ and $\EE, \FF$, respectively. The Kraus operators of the second-order superswitch are then defined as
\begin{align}
	K^{(2)}_{ijkl}&=K^{(1)}_{ij} \kk^{(1)}_{kl} \otimes \ketbra{0}_3+ \kk^{(1)}_{kl} K^{(1)}_{ij} \otimes \ketbra{1}_3 \notag \\
	&= \frac{1}{2} \left\{K^{(1)}_{ij},\kk^{(1)}_{kl}\right\}\otimes \Id+ \frac{1}{2}\left[\kk^{(1)}_{kl},K^{(1)}_{ij}\right]\otimes Z,
\end{align}
\begin{figure}[!t]
	\centering	\includegraphics[width=0.7\linewidth]{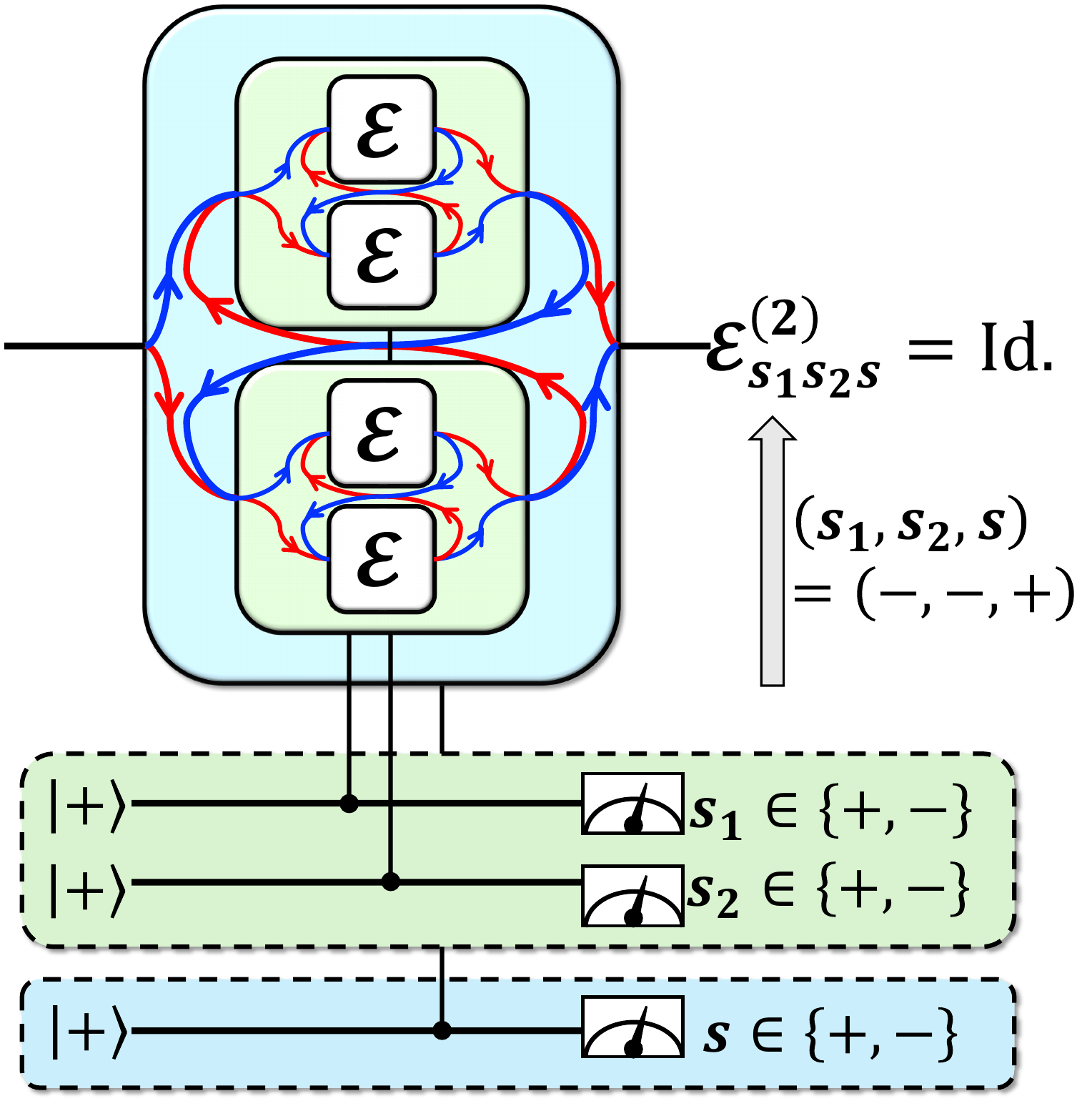}
	\caption{A visual representation of the second-order superswitch.}
	\label{Fig: first-order superswitch}	
\end{figure} 

We readily obtain the second order superswitch as 
\begin{align}
	\Sw^{(2)}_{\omega_c}(\E,\F,\EE,\FF)(\rho)=\sum_{i,j,k,l} K_{ijkl}^{(2)}(\rho\otimes \omega_c) \left(K_{ijkl}^{(2)}\right)^\dagger \,,
\end{align}
where the ancilla state $\omega_c$ controls the order of channels $\E, \F$ and $\EE, \FF$ inside the two first-order superswitches respectively, as well as the ordering of the two superswitches themselves. A schematic representation is shown in Fig. \ref{Fig: first-order superswitch}. Expanding the Kraus operators we find
\begin{align}
	K_{ijkl}^{(2)}=&E_i F_j \ee_k \ff_l \otimes \ketbra{0}\otimes \ketbra{0}\otimes \ketbra{0} \notag \\
	+& E_i  F_j \ff_l \ee_k  \otimes \ketbra{0}\otimes \ketbra{1}\otimes \ketbra{0} \notag \\
	+& F_j E_i  \ee_k \ff_l \otimes \ketbra{1}\otimes \ketbra{0}\otimes \ketbra{0} \notag \\
	+ & F_j E_i \ff_l  \ee_k  \otimes \ketbra{1}\otimes \ketbra{1}\otimes \ketbra{0} \notag \\
	+&\ee_k \ff_l  E_i F_j \otimes \ketbra{0}\otimes \ketbra{0}\otimes \ketbra{1} \notag \\
	+& \ff_l \ee_k E_i  F_j  \otimes \ketbra{0}\otimes \ketbra{1}\otimes \ketbra{1} \notag \\
	+& \ee_k \ff_l  F_j E_i  \otimes \ketbra{1}\otimes \ketbra{0}\otimes \ketbra{1} \notag \\
	+& \ff_l  \ee_k  F_j E_i   \otimes \ketbra{1}\otimes \ketbra{1}\otimes \ketbra{1} \,, \label{eq: kraus 1st order superswitch}
\end{align} 
where we have dropped the subscripts from the ancilla qubits. This expression can also be written in terms of commutators and anticommutators as
\begin{align}
	K_{ijkl}^{(2)} &= \notag \\
	  \frac{1}{8} &\sum_{s_1, s_2, s} \Big[\big[E_i,F_j\big]_{s_1},\big[\ee_k,\ff_l\big]_{s_2}\Big]_s \otimes P_{s_1} \otimes P_{s_2} \otimes P_s \notag \\
	\frac{1}{8}&\Big( \acom{\acom{E_i}{F_j}}{\acom{\ee_k}{\ff_l}} \otimes \Id \otimes \Id \otimes \Id \notag \\
	&+ \acom{\com{E_i}{F_j}}{\acom{\ee_k}{\ff_l}} \otimes Z\otimes \Id \otimes \Id 
	\notag \\
	&+\acom{\acom{E_i}{F_j}}{\com{\ee_k}{\ff_l}} \otimes \Id \otimes Z \otimes \Id \notag \\
	&+  \acom{\com{E_i}{F_j}}{\com{\ee_k}{\ff_l}} \otimes Z \otimes Z \otimes \Id \notag \\
	& + \com{\acom{E_i}{F_j}}{\acom{\ee_k}{\ff_l}} \otimes \Id \otimes \Id \otimes Z \notag \\ &+\com{\com{E_i}{F_j}}{\acom{\ee_k}{\ff_l}}\otimes Z\otimes \Id \otimes Z
	\notag \\
	& +\,\,\com{\acom{E_i}{F_j}}{\com{\ee_k}{\ff_l}} \otimes \Id \otimes Z \otimes Z \notag \\
	&+\,\, \com{\com{E_i}{F_j}}{\com{\ee_k}{\ff_l}}\otimes Z \otimes Z \otimes Z\Big)\,,
\end{align}
where we have introduced the notation $[A,B]_{+} \equiv \{A,B\}=AB+BA$ and $[A,B]_{-} \equiv [A,B]=AB-BA$ for the anticommutator and commutator, and $P_{+} =\Id$, $P_{-} = Z$. 
We now assume that for any pair of indices $i$ and $j$, the commutator $[X_i,Y_j]$ and anticommutator $\{X_i,Y_j\}$ of a pair of Kraus operators $X_i, Y_j$ are not simultaneously nonzero, which holds for Pauli channels, we then obtain the second-order superswitch,
\begin{align}
	&\Sw^{(2)}_{\omega_c}(\E,\F,\EE,\FF)(\rho)=\sum_{i,j,k,l} K_{ijkl}(\rho\otimes\omega_c)K_{ijkl}^\dagger \notag \\
	&= \frac{1}{64} \sum_{i,j,k,l} \sum_{s_1, s_2, s} \notag \\
	&\Big[\big[E_i,F_j\big]_{s_1},\big[\ee_k,\ff_l\big]_{s_2}\Big]_s \rho \Big[\big[E_i,F_j\big]_{s_1},\big[\ee_k,\ff_l\big]_{s_2}\Big]_s^\dagger \notag \\
	& \hspace{20mm}\otimes ( P_{s_1} \otimes P_{s_2} \otimes P_s) \omega_c ( P_{s_1} \otimes P_{s_2} \otimes P_s)^\dagger \,,
	\label{eq: 1st order superswitch}
\end{align}
where the indices $i,j,k,l \in [1,\ldots,4]$ and $s_1,s_2,s =\pm$.

We further choose the ancilla state to be a product state $\omega_c = \omega_1 \otimes \omega_2 \otimes \omega$, with $\omega_1$ and $\omega_2$ controlling the ordering of the channels $\E, \F$ and $\EE, \FF$, respectively, in the first-order superswitches and $\omega$ controlling their ordering. Letting  $\omega_1=\omega_2=\omega=\ketbra{+}$ and defining channels $\E_{ijk}^{(2)}$ as, e.g.
\begin{align}
	&\symsc{\E^{(2)}}{--+}(\E,\F,\EE,\FF)(\rho) = \notag \\
	 &\frac{1}{\symsc{q}{--+}} \frac{1}{64}\sum_{i,j,k,l}\acom{\com{E_i}{F_j}}{\com{\ee_k}{\ff_l}}\rho \acom{\com{E_i}{F_j}}{\com{\ee_k}{\ff_l}}^\dagger\,,
\end{align}
where
\begin{align}
	&\symsc{q}{--+}^{(2)} = \notag \\
	&	\tr\Bigg(\frac{1}{64 }\sum_{i,j,k,l}\acom{\com{E_i}{F_j}}{\com{\ee_k}{\ff_l}}\rho \acom{\com{E_i}{F_j}}{\com{\ee_k}{\ff_l}}^\dagger\Bigg),
\end{align}
are the associated probabilities of ocurrence. The second-order superswitch takes the form
\begin{align}
	\Sw^{(2)}&(\E,\F,\EE,\FF)(\rho)= \notag \\
	&\sum_{s_1,s_2,s = \pm} q_{s_1 s_2 s }^{(2)} \E^{(2)}_{s_1 s_2 s} (\E, \F, \EE, \FF) \otimes \ketbra{s_1 s_2 s}\notag \\
	= \,\,\,\,\,&\symsc{q}{+++}^{(2)} \symsc{\E^{(2)}}{+++}(\E,\F,\EE,\FF)(\rho) \otimes\ketbra{+++} \notag \\
	+&\symsc{q}{-++}^{(2)} \symsc{\E^{(2)}}{-++}(\E,\F,\EE,\FF)(\rho)\otimes \ketbra{-++}	\notag \\
	+&\symsc{q}{+-+}^{(2)} \symsc{\E^{(2)}}{+-+}(\E,\F,\EE,\FF)(\rho)\otimes \ketbra{+-+}	\notag \\
	+&\symsc{q}{--+}^{(2)} \symsc{\E^{(2)}}{--+}(\E,\F,\EE,\FF)(\rho)\otimes \ketbra{--+}	\notag \\
	+&\symsc{q}{++-}^{(2)} \symsc{\E^{(2)}}{++-}(\E,\F,\EE,\FF)(\rho)\otimes\ketbra{++-} \notag \\
	+&\symsc{q}{-+-}^{(2)} \symsc{\E^{(2)}}{-+-}(\E,\F,\EE,\FF)(\rho)\otimes \ketbra{-+-}	\notag \\
	+&\symsc{q}{+--}^{(2)} \symsc{\E^{(2)}}{+--}(\E,\F,\EE,\FF)(\rho)\otimes\ketbra{+--}	 \notag \\
	+& \symsc{q}{---}^{(2)} \symsc{\E^{(2)}}{---}(\E,\F,\EE,\FF)(\rho) \otimes \ketbra{---} \,.
\end{align}
We let $\E=\EE=\F=\FF=\P_{\vec{p}}=\sum_i p_i \sigma_i \rho \sigma_i^\dagger$, that is, instances of the same channel are combined in the first-order superswitch, and we list the general expressions. For notational brevity, we will denote a Pauli channel $\E=\P_{\vec{p}}=p_0 \rho+ p_1X\rho X +p_2 Y\rho Y+p_2 Z \rho Z$ with the last three components of its probability vector $\vec{p}$, since, due to the normalisation condition the first follows from the other three, i.e. $p_0=1-p_1-p_2-p_3$, and we will write $\E = \{p_1,p_2,p_3\}$. For instance, we will write  $\D_p = \{\frac{p}{4},\frac{p}{4},\frac{p}{4}\}$ to denote a depolarisation channel. We this notation, we can list all the eight $\E_{ijk}^{(2)}$ channels of the second-order superswitch:
\begin{align}
	\symsc{\E^{(2)}}{+++}=& \frac{4 p_0  (p_0^2 + p_1^2 + p_2^2 + p_3^2)}{\q{+++}^{(2)}} \{p_1,p_2,p_3\} \,, \notag \\
	\symsc{\E^{(2)}}{-++}=&\frac{2  (p_0^2 + p_1^2 + p_2^2 + p_3^2)}{\q{-++}^{(2)}} \{p_2 p_3, p_1 p_3, p_1 p_2 \}\,, \notag\\
	\symsc{\E^{(2)}}{+-+}=&\symsc{\E^{(2)}}{-++} \,, \notag \\
	\symsc{\E^{(2)}}{--+}=& \{0,0,0 \} = \id\,, \notag\\
	\symsc{\E^{(2)}}{++-}=& \frac{8 p_0^2}{\q{++-}^{(2)}} \{p_2 p_3 ,  p_1 p_3 , p_1 p_2 \} \notag\\
	\symsc{\E^{(2)}}{-+-}=&\frac{4 p_0}{\q{-+-}^{(2)}} \{p_1 (p_2^2 + p_3^2) ,  p_2 (p_1^2 + p_3^2) ,  p_3 (p_1^2 + p_2^2)\}\,, \notag\\
	\symsc{\E^{(2)}}{+--}=&\symsc{\E^{(2)}}{-+-}\,, \notag \\
	\symsc{\E^{(2)}}{---}=&\frac{8p_1 p_2 p_3}{\q{---}^{(2)}} \{p_1 , p_2,  p_3\} \,, \label{eq: channels 1st}
\end{align}
with the respective probabilities of occurrence 
\begin{align}
	\q{+++}^{(2)}&= p_0^4 + 4 p_0^3 (p_1 + p_2 + p_3) + 6 p_0^2 (p_1^2 + p_2^2 + p_3^2) \notag \\
	&\hspace{-1mm}  + 4 p_0 (p_1 + p_2 + p_3) (p_1^2 + p_2^2 + p_3^2) + (p_1^2 + p_2^2 + p_3^2)^2,\notag  \\
	\q{-++}^{(2)}&=  12 p_0 p_1 p_2 p_3 + \notag \\
	&\hspace{5mm} + 2(p_0^2 +p_1^2 + p_2^2 + p_3^2) (p_1 p_2 +p_2 p_3 + p_3 p_1) , \notag\\
	\q{+-+}^{(2)}& = \q{-++}^{(2)}, \notag \\
	\q{--+}^{(2)}&= 4 \left( p_1^2 p_2^2 + p_2^2 p_3^2 +  p_3^2 p_1^2 \right), \notag \\
	\q{++-}^{(2)}&= 8 p_0^2 (p_1 p_2 +p_2 p_3 + p_3 p_1), \notag \\
	\q{-+-}^{(2)}&= 4 p_0 \left( p_1^2 (p_2 + p_3) + p_2^2  (p_3 + p_1) + p_3^2 (p_1 + p_2) \right), \notag \\
	\q{+--}^{(2)}&= \q{-+-}^{(2)}, \notag \\
	\q{---}^{(2)}&= 8 p_1 p_2 p_3 (p_1 + p_2 + p_3) \,. \label{eq: probs 1st}
\end{align}
Note that the $\symsc{\E^{(2)}}{--+}$ channel is the identity channel, $\id$, which is the crucial feature that enables our results.

We remark that in Ref.\@ \cite{das_quantum_2022} the concept of switch of switches was introduced, similar to the case we consider here but in a restricted scenario where there is only a single control ancilla for the inside switches, i.e.~their ordering is `synchronised'. Here we use the general case which is crucial for our purposes, as the synchronised switches in Ref.\@ \cite{das_quantum_2022} do not exhibit the same properties. The reason for this is that the effect of forcing synchronisation leads to mixing of terms \cite{kechrimparis_enhancing_2024}, and thus the possibility to get the identity channel is in general removed.

\section{Geometric picture of Pauli channels in $d=2$}
\label{app: geometric picture of Pauli channels in d=2}
A completely positive and trace preserving map (CPTP) acting on states in dimension $d=2$ can be written in the general form
\begin{align}
	\tilde{\E}(\rho) = \U  \circ  \E( \rho ) \circ \mathcal{V}\,,
\end{align}
where $\E$ can be seen as acting on the Bloch vector of the state as
\begin{align}
	\E(\rho)= \frac{1}{2}\left( \Id +(M\vec{r} + \vec{t})\cdot \vec{\sigma} \right)\,,
\end{align}
with $M=\text{diag}(\lambda_1,\lambda_2,\lambda_3)$ and $\vec{t}=(t_1,t_2,t_3)$. The $\lambda_i$ satisfy $(\lambda_1 \pm \lambda_2)^2 \leq (1\pm \lambda_3)^2$, along with permutations of the indices. The conditions on the elements of $M$ and $\vec{t}$ for a map to be CPTP are given in Ref.\@ \cite{ruskai_analysis_2002}. Geometrically, a channel will map the Bloch ball into a displaced ellipsoid inside it. However, not all ellipsoids inside the Bloch ball correspond to CPTP maps. 

The case $\vec{t}=0$ corresponds to unital channels and in terms of the parametrisation we have been using so far, i.e. $\E(\rho)=\sum_i p_i \sigma_i \rho \sigma _i$, we have
\begin{align}
	\lambda_i =1-2p_j -2p_k \,,
\end{align}
with $i,j,k \in \{1,2,3\}$ and $i\neq j\neq k$. 
\begin{figure}[!t]
	\includegraphics[width=0.9\columnwidth]{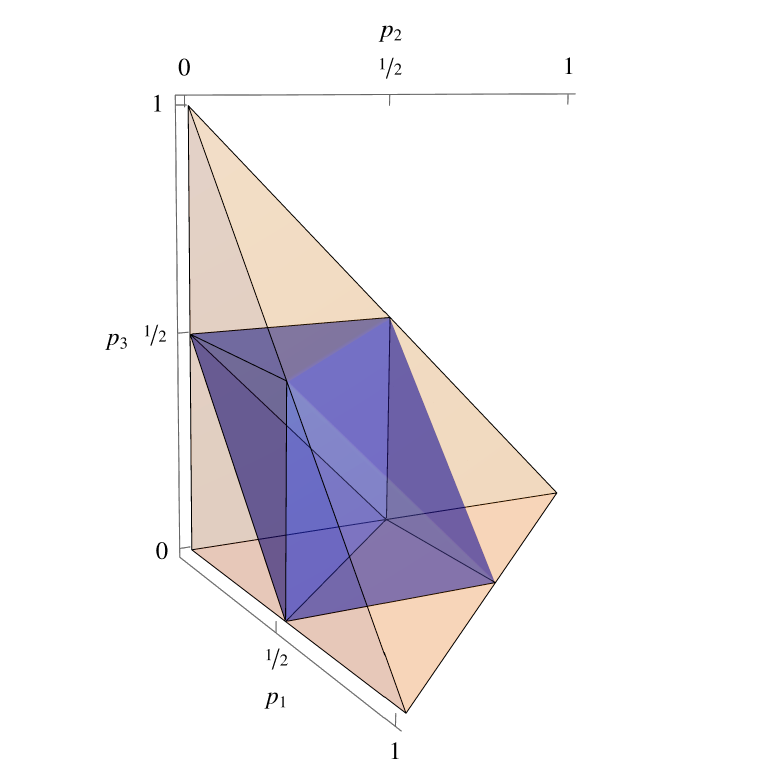}
	\caption{The set of unital channels, corresponding to the tetrahedron in yellow. The octahedron in blue corresponds to the set of entanglement-breaking channels.  \label{fig: unital channels}}
\end{figure}

The Kraus rank of a channel is the minimum number of Kraus operators among all Kraus decompositions. Kraus rank 1 maps are the $\id$ channel and the three unitary channels $\U_{X},\U_{Y},\U_{Z}$, which correspond to the vertices of the tetrahedron. Similarly Kraus 2 operations are on the edges, while Kraus 3 are on the faces. Any map inside the tetrahedron necessarily has Kraus rank 4.

A Pauli channel is entanglement breaking if the following conditions are satisfied \cite{davalos_divisibility_2019,horodecki_entanglement_2003}
\begin{align}
	\lambda_1+\lambda_2 +\lambda_3 &\leq 1\,, \notag \\
	\lambda_i-\lambda_j -\lambda_k &\leq 1\,, \,\, \forall \, i\neq j\neq k \,.
\end{align}
 It is straightforward to check that in terms of the $p_i$ parametrisation of unital channels, these conditions are equivalent to $p_i \leq\frac{1}{2}\,, \forall i $. These define an octahedron inside the tetrahedron of unital channels.

From the discussion in the main text, including Result 1, at the faces of the tetrahedron lie channels that can be purified by employing the quantum switch. Channels of Kraus rank 4 lie strictly inside the tetrahedron and can only be purified with the second-order superswitch. Thus, with the quantum switch only a particular subset of entanglement-breaking channels (i.e. those at the intersection of the surfaces of octahedron and tetrahedron) can be distilled. On the other hand, the second-order superswitch enables distillation of all entanglement-breaking channels. The tetrahedron of unital channels, as well as the octahedron corresponding to the set of entanglement-breaking channels are shown in Fig.\@ \ref{fig: unital channels}.

\section{Proof of Results 1 and 2}
\label{app: proof of results 1 and 2}
Result 1 follows by noting that in the expressions for the second-order superswitch, Eqs.\@ \eqref{eq: channels 1st} and \eqref{eq: probs 1st}, the outcome $`- - +'$ after measurements in the ancillas: (i) leads to the identify channel $\id$, for \emph{any} input Pauli channel $\E_{\vec{p}}$, and (ii) this probability is nonzero, i.e.  $\q{--+}>0$, for all channels except for the trivial case of $\E_{\vec{p}}=\id$ as well as channels with two of the $p_i$ with $i=1,2,3$ equal to zero. However, the latter case can be mapped to the case with only one of the $p_i$ being zero.
Specifically, applying one of the Pauli matrices $\sigma_i$ with the same index $i$ as the zero $p_i$ to the channel before inputting in the superswitch, maps each channel to a channel with only one zero $p_i$, thus achieving nonzero probability $\q{--+}^{(2)}$ according to Result 1, and thus all channels are covered. For example, by applying either a unitary $Y$ or $Z$ to the channel $(1-p)\rho + p X\rho X$, maps it to either $(1-p)Y\rho Y + p Z\rho Z$ or $(1-p)Z\rho Z + p Y\rho Y$, respectively. Thus, all channels are covered by Result 1.

The fact that the second-order superswitch is a switch of two switches is what enables this phenomenon. Specifically, this follows directly from the equations of the quantum switch \cite{chiribella_quantum_2013, chiribella_indefinite_2021}, 
in the case of two channels $\E=\F= p_0\rho+p_1 X\rho X +p_2 Y\rho Y+ p_3 Z\rho Z$ 
with $\sum_i p_i  =1$, which reduce to $S_\omega (\rho) = \q{+} \scale{C}{+}(\rho) \otimes \scale{\omega}{+} + \q{-} \scale{C}{-} (\rho) \otimes \scale{\omega}{-}$. The channels $C_+, C_-$ \cite{chiribella_indefinite_2021} are defined as
\begin{align}
	\scale{C}{+}(\rho) &=\frac{1}{\q{+}}\Big(\Big(\sum_{i=0}^4 p_i^2\Big)\rho+2p_0\Big(\sum_{i=1}^4 p_i \sigma_i \rho \sigma_i\Big) \Big)\,,\notag \\
	\scale{C}{-}(\rho) &= \frac{2}{\q{-}} \Big(p_1 p_2 \sigma_3\rho \sigma_3+ p_2 p_3 \sigma_1\rho \sigma_1 +  p_3 p_1 \sigma_2 \rho \sigma_2\Big)\,, \label{eq:switch plus and minus channels}
\end{align}
with $\q{-} = 2(p_1 p_2 + p_2 p_3 +p_3 p_1)$ and $\q{+} = 1-\q{-}$, $\scale{\omega}{+} =\omega$ and $\scale{\omega}{-} = Z\omega Z$. The choice $\omega = \proj{+}$ leads to $\scale{\omega}{\pm} = \proj{\pm}$ and thus by measuring the ancilla on the same basis, the two channels can be fully separated.

It readily follows that: (i) for any Pauli channel, two copies of which are combined in the quantum switch, the minus outcome after measurement on the ancilla qubit leads to a Pauli channel with $p_0=0$, and (ii) for any Pauli channel with $p_0=0$, two copies of which are combined in the quantum switch, the plus outcome after measurement on the ancilla qubit leads to the identity channel $\id$. Then, a way to satisfy both conditions at the same time is to construct the second-order superswitch: whenever measurements on the ancillas of the two inside switches give minus outcomes and measurement on the ancilla controlling the order of the switches gives a plus outcome, the above scenario occurs and we thus obtain a noiseless copy.

Result 2 follows easily from the functional form of the distillation rate, 
\begin{align}
	\R^{(2)} = \q{--+}^{(2)} = 4 \left( p_1^2 p_2^2 + p_2^2 p_3^2 +  p_3^2 p_1^2 \right)\,,
\end{align}
and the observation that as all terms are positive, given two channels $\E=\P_{\vec{p}}$ and $\F=\P_{\vec{q}}$ with probability vectors $\vec{p}, \vec{q}$ such that $p_i \geq q_i$ for all $i\in\{1,2,3\}$, then we necessarily have $\R^{(2)}_{\E} \geq \R^{(2)}_{\F}$. The fact that $p_i \geq q_i$ for all $i\in\{1,2,3\}$ implies that the channel $\E$ is farther away from the identity channel, $\id$. 

To make this more mathematically precise, we compare the difference of their trace distances by defining the quantity
\begin{align}
	d(\E,\F,\rho)=T(\rho,\E(\rho))-T(\rho,\F(\rho)) \,,
\end{align}
where $T$ denotes the trace distance,
\begin{align}
	T(\rho, \sigma) =\frac{\norm{\rho-\sigma}_1}{2} = \frac{1}{2} \tr(\sqrt{(\rho-\sigma)^\dagger (\rho-\sigma)}) \,.
\end{align}
Note that the quantity $d(\E,\F,\rho)$ is not a distance as it can be negative and is not symmetric in the channels. 
As for two qubit states $\rho, \sigma$, the trace distance is related to the Euclidean distance $d_e$ of their Bloch vectors $\vec{r}, \vec{s}$, respectively, and we have that
\begin{align}
	T(\rho, \sigma) &= \frac{d_e(\vec{r},\vec{s})}{2}  \notag \\
	&= \frac{\sqrt{(r_1-s_1)^2+(r_2-s_2)^2+(r_3-s_3)^2}}{2}\,.
\end{align}
The action of a quantum channel $\N$ in dimension $d=2$ is \cite{ruskai_analysis_2002}
\begin{align}
	\N(\rho) = \frac{1}{2} \left(\Id + (M \vec{r}+\vec{t})\cdot\vec{\sigma}\right)\,.
\end{align}
\begin{figure}[!t]
	\includegraphics[width=0.9\columnwidth]{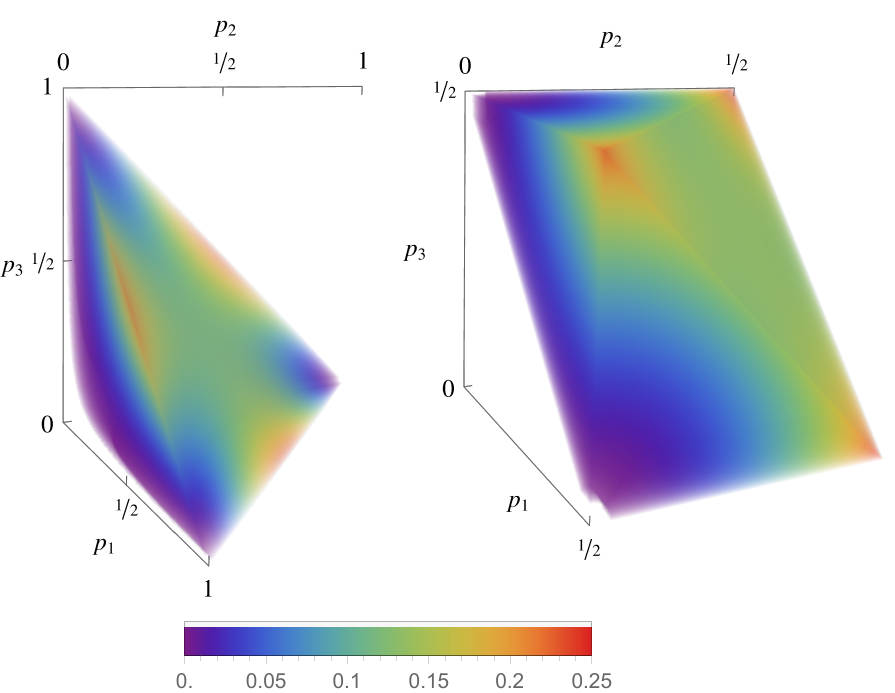}
	\caption{Visualisation of the distillation rate $\R^{(2)}$ for (i) all unital channels (left) and (ii) the set of entanglement-breaking channels (right).  \label{fig: com rate unital and EB channels}}
\end{figure}
 Restricting to the case of a Pauli channel, the Bloch vector is transformed as
\begin{align}
	(r_1 ,r_2, r_3) \rightarrow \Big( &(1-2p_2-2p_3)r_1,(1-2p_3-2p_1)r_2, \notag \\
	 &\hspace{15mm}(1-2p_1-2p_2)r_3\Big).
\end{align}
 Then, for two Pauli channels $\E=\P_{\vec{p}}$ and $\F=\P_{\vec{q}}$ such that $p_i \geq q_i$ for all $i\in\{1,2,3\}$, we have
\begin{align}
		d(\E,\F,\rho)&=\sqrt{r_1^2 (p_2+p_3)^2 +r_2^2 (p_3+p_1)^2 +r_3^2 (p_1+p_2)^2}\notag \\ &-\sqrt{r_1^2 (q_2+q_3)^2 +r_2^2 (q_3+q_1)^2 +r_3^2 (q_1+q_2)^2} \notag \\
		&\hspace{0mm} \geq 0\,,
\end{align}
which can be restated as the statement that $\forall \rho$
\begin{align}
		\norm{\rho-\E(\rho)}_1\geq \norm{\rho-\F(\rho)}_1\,,
\end{align}
which proves the claim. The distillation rates for Pauli channels in the interior of the tetrahedron are visualised in Fig.\@ \ref{fig: com rate unital and EB channels}.

\section{Distillation rates for $n$-order superswitches}
\label{app: distillation rate for n-order superswitches}
As discussed in Appendix \ref{app: n-order superswitches}, for any Pauli channel $\E_{\vec{p}}$, four copies of which are combined in the second-order superswitch there is a nonzero probability $\q{--+}^{(2)}>0$, that the outcome `$- - +$' is obtained. By definition, the distillation rate $\R^{(2)}$ is the probability that the outcome that leads to the identity channel is obtained, i.e. `$- - +$', and thus for the second-order superswitch is
\begin{align}
	\R^{(2)} = \q{--+} = 4 \left( p_1^2 p_2^2 + p_2^2 p_3^2 +  p_3^2 p_1^2 \right)\,.
\end{align}

The observation regarding the second-order superswitch and the ability to lead to the identity channel with nonzero probability for any channel holds for higher-order superswitches as well. Specifically, at order $n$ there are $2^{n}-1$ ancilla qubits. The last ancilla controls the order of the inside two $(n-1)$-order superswitches, while the second and third from the end control the ordering of lower-order superswitches inside these two $(n-1)$-order superswitches. 
It follows that whatever the previous outcomes $o_1, o_2, \ldots$ on the rest of the ancilla qubits, the resulting channel is the identity channel, with some probability  $0\leq \q{o_1 \cdots - - +}^{(n)} \leq1$. We note that some sequences of outcomes may have probability of occurrence 0.
This argument ignores the possibility of unitary channels which can arise on the face of the tetrahedron and thus applies only in the interior. The complete derivation of the distillation rates that include the faces is given in Appendix \ref{app: Recurrence Relations and Asymptotic Distillation Rate}.  With this in mind, the distillation rate $\R^{(n)}$ of an $n$-order superswitch is then
\begin{align}
	\R^{(n)} = \sum_{o_1 ,\ldots=\pm } \Pr{o_1, \cdots, o_{2^{n}-4}, - , - , +}
\end{align}

For the depolarisation channel, the distillation rates of up to order $n=5$ are:
\begin{align}
	\R^{(2)}&=\frac{3 p^4}{64} \,, \notag \\
	\R^{(3)}&= \frac{3}{64} (-2 + p)^4 p^4 + \frac{9 p^8}{4096}\,, \notag \\
	\R^{(4)}&=12 p^4 - 72 p^5 + 210 p^6 - 390 p^7 + \frac{4095 p^8}{8} - \frac{999 p^9}{2} \notag \\
	&+ \frac{11877 p^{10}}{32} - \frac{6759 p^{11}}{32} + \frac{746673 p^{12}}{8192} - \frac{118833 p^{13}}{4096}\notag \\ 
	&+ \frac{105651 p^{14}}{16384} - \frac{14673 p^{15}}{16384} + \frac{981201 p^{16}}{16777216} \,, \label{eq: distil rate 1}
\end{align}
and
\begin{align}
	\R^{(5)}&= 192 p^4 - 2688 p^5 + 19488 p^6 - 96096 p^7 + 359052 p^8 \notag \\
	&- 1076304 p^9 + 2679264 p^{10} - 5663592 p^{11} \notag \\
	&+ \frac{660468723 p^{12}}{64} - \frac{524016063 p^{13}}{32} + \frac{2916648909 p^{14}}{128} \notag \\
	&- \frac{3575251239 p^{15}}{128} + \frac{247771513233 p^{16}}{8192} \notag \\
	&- \frac{29664177201 p^{17}}{1024} + \frac{402111190275 p^{18}}{16384} \notag \\
	&- \frac{301008655485 p^{19}}{16384} + \frac{25426529527977 p^{20}}{2097152} \notag \\
	&- \frac{7371155519931 p^{21}}{1048576} + \frac{14948591861835 p^{22}}{4194304} \notag \\
	&- \frac{6585748985817 p^{23}}{4194304} + \frac{40009926738525 p^{24}}{67108864} \notag \\
	&- \frac{3238417329813 p^{25}}{16777216} + \frac{14102865567135 p^{26}}{268435456} \notag \\
	&- \frac{3165750102405 p^{27}}{268435456} + \frac{146071342742253 p^{28}}{68719476736} \notag \\
	&- \frac{10160062539117 p^{29}}{34359738368} + \frac{4095741366951 p^{30}}{137438953472} \notag \\
	&- \frac{266117803725 p^{31}}{137438953472} + \frac{17141597128353 p^{32}}{281474976710656}
	\,. \label{eq: distil rate 2}
	\end{align}

\section{The Choi-Jamiołkowski isomorphism and entanglement sharing} 
\label{app: The Choi-Jamiołkowski isomorphism and entanglement sharing}
Our results can be directly translated to an entanglement sharing scenario. In this setting, a party prepares some entangled state, keeps one part of that state, and sends the other to the second party through some channel $\E$. Clearly, if the channel is entanglement-breaking, this task cannot be performed. In addition, unless the channel is the identity, $\id$, or a unitary channel, sharing a maximally entangled state is also impossible. Our main results, however, show that this is possible given access to at least the second-order superswitch.
This follows from the Choi-Jamiołkowski isomorphism \cite{choi_completely_1975,jamiolkowski_linear_1972,jiang_channelstate_2013}, which establishes a correspondence between bipartite states and channels.  Specifically, a map between channels and bipartite states is established through the equation
\begin{align}
	\rho = (\id \otimes \E) [\proj{\Psi}] \,,
\end{align}
where $\proj{\Psi}$ is a maximally entangled state.
Pauli channels are mapped to Bell-diagonal states, which are mixtures of the four entangled states. Specifically, a Pauli channel $\P$ is mapped to the Bell diagonal state
\begin{align}
	\rho_B &= p_0 \proj{\Phi^+}+p_1 \proj{\Psi^+} \notag \\
	&+p_2 \proj{\Psi^-}+p_3 \proj{\Phi^-} \,, \label{eq:Bell diagonal}
\end{align}
where 
\begin{align}
	\ket{\Phi^+} &=\frac{1}{\sqrt{2}}(\ket{00}+\ket{11})\,, \,\, \ket{\Phi^-} =\frac{1}{\sqrt{2}}(\ket{00}-\ket{11})\,, \notag \\
	\ket{\Psi^+} &=\frac{1}{\sqrt{2}}(\ket{01}+\ket{10})\,, \,\, \ket{\Psi^-} =\frac{1}{\sqrt{2}}(\ket{01}-\ket{10})\,, \notag \\
\end{align}
are the four Bell states. We note that a Bell diagonal state is separable if 
\begin{align}
	p_{\max} = \max_i  \{p_i \} \leq \nicefrac{1}{2}\,.
\end{align}

The depolarisation channel, defined with parameters $p_1=p_2=p_3 = \nicefrac{p}{4}$ is mapped to 
\begin{align}
	\rho= (1-p)  \proj{\Phi^+}+p \frac{\Id_4}{4}\,,
\end{align}
which is a Werner state \cite{werner_quantum_1989}
\begin{align}
	\rho_W= \lambda  \proj{\Psi^-}+(1-\lambda) \frac{\Id_4}{4}\,,
\end{align}
up to a unitary transformation. It is known that Werner states are entangled whenever $\lambda>\nicefrac{1}{3}$.

Through the Choi-Jamiołkowski isomorphism, we show that results on channels are translated to results on bipartite states through the quantum switch, which gives
\begin{align}
	\tilde{S}_\omega (\F,\F) =& \frac{1}{4} \sum_{i,j} \Big(\{F_i,F_j\}\rho\{F_i,F_j\}^\dag \otimes \omega \notag \\
	&+[F_i,F_j]\rho[F_i,F_j]^\dag \otimes Z\omega Z  \Big) \,,  \label{eq: state quantum switch commutators}
\end{align}
where the channel $\F$ is acting on bipartite states $\rho_{AB} $ and is given by $\F=\Id\otimes \E$, where $\E$ is a channel acting on system $B$. Then, the Kraus operators are explicitly $F_i=\Id\otimes E_i$. If $\E$ is a Pauli channel, we find that the result of applying the quantum switch is
\begin{align}
	\tilde{S}_\omega (\E, \E) = q_+ \rho_{AB}^{(+)}  \otimes \omega_+ + q_- \rho_{AB}^{(-)}  \otimes \omega_- \,, \label{eq: state quantum switch}
\end{align}
where the bipartite states $\rho_{AB}^{(+)} $ and $\rho_{AB}^{(-)} $ are defined as
\begin{align}
	\rho_{AB}^{(+)}  &= \Big(p_0^2+p_1^2+p_2^2+p_3^2)\proj{\Phi^+}+2p_0(p_1 \proj{\Psi^+} \notag \\
	&\hspace{20mm}+p_2 \proj{\Psi^-} +p_3 \proj{\Phi^-}\Big)\big/ q_+ \notag \\
	\rho_{AB}^{(-)}  &= \Big(2 p_1 p_2 \proj{\Phi^-} + 2p_2 p_3 \proj{\Psi^+} \notag \\
		& \hspace{20mm}+ 2 p_3 p_1 \proj{\Psi^-}) \Big)\Big/ q_- \label{eq:switch plus and minus states}
\end{align}
with
\begin{align}
	q_- = 2(p_1 p_2 + p_2 p_3 +p_3 p_1) \,,\, \, \, q_+ = 1-q_-\,, 
\end{align}
and $\omega_+ =\omega$ and $\omega_- = Z\omega Z$. The choice $\omega = \proj{+}$ leads to $\omega_{\pm} = \proj{\pm}$, and thus a measurement in the $\ket{\pm}$ basis can separate the two states completely. A similar picture holds for the higher-order superswitches. For instance, the second-order superswitch follows directly by applying Eq.\@ \eqref{eq:Bell diagonal} to the expressions Eqs.\@ \eqref{eq: channels 1st}-\eqref{eq: probs 1st}.

\section{Twirling channels and application to superswitches}
\label{app: Twirling channels and application to superswitches}
Given a channel $\E$, twirling  is a sort of averaging with random applications of unitaries over the Haar measure \cite{nielsen_simple_2002,gross_evenly_2007}. It can be seen as a supermap that maps any channel to the depolarising channel. Specifically, it is defined through the equation
\begin{align}
	\T[\E](\rho)&=\int  \mathrm{d}U\, U^\dagger \E(U\rho U^\dagger) U \notag \\
    &= (1-\eta_\E)\rho+\eta_\E \frac{\Id}{2} \equiv \D_{\eta_\E}(\rho)
\end{align}
in the case with $d=2$. If the channel is expressed with Kraus operators as $\E(\rho)=\sum K_i\rho K_i^\dagger$, then the depolarising parameter is obtained as
\begin{align}
	1-\eta_\E =\frac{\sum_i\abs{\tr K_i^2}-1}{3}\,.
\end{align}
Note that in the special case of a Pauli channel $\P$, the last equation reduces to
\begin{align}
	1-\eta_\P =\frac{4p_0-1}{3}\,.
\end{align}

A \emph{unitary 2-design} \cite{gross_evenly_2007} is a finite set of unitary matrices $\mathcal{UD}=\{U_i\}_{i=1,\ldots,N}$ that can perform twirling. That is,
\begin{align}
	\frac{1}{N}\sum_i U_i^\dagger \E(U_i\rho U_i^\dagger) U_i \equiv \T_\E(\rho)=(1-\eta_\E)\rho+\eta_\E \frac{\Id}{2}\,.
\end{align}
In dimension $d=2$, the Clifford group is an example of a unitary 2-design. It consists of 24 elements but there exist unitary 2-designs with fewer elements, with 10 being the lower bound. Moreover, it is known that for the restricted class of Pauli channels, three unitaries are sufficient to perform twirling, making this protocol practical \cite{kechrimparis_measurementprotected_2019}.
The above machinery is important for our purposes as it allows us to turn any channel into a depolarisation one and thus the statements made for Pauli channels, apply implicitly to non-unital channels as well.  For a single use of a channel, and a unitary-2 design with $N$ elements, in practice one can perform twirling by:
\begin{enumerate}
	\item producing $N$ copies of the state,
	\item $\forall i$ applying a different $U_i$ to each copy, 
	\item sending each $U_i \rho U_i ^\dagger$ through a copy of the channel $\E$,
	\item applying a $U_i^\dagger$ to each copy with the same index,
	\item using each copy for whatever purpose (e.g. measuring) and
	\item averaging over outcomes over all copies, thus obtaining $\T[\E](\rho)= \D_{\eta}(\rho)$.
\end{enumerate} 

However, as the superswitches combine multiple copies of the channel, this procedure needs to be adapted. We describe the modified implementation for the case of the quantum switch, which combines two copies of the channel, and the generalisation to higher-order superswitches is straightforward. Let us denote by $\E_i$ the channels $\E_i = \U_i^\dagger \circ \E \circ \U_i $, where $\U_i(\rho) = U_i \rho U_i^\dagger$ denotes a unitary channel and $\U_i^\dagger$ its adjoint. Each channel $\E_i$ has Kraus operators $E_{ik} = U_i^\dagger E_k U_i$, where $E_k$ are the Kraus operators of $\E$. Then, it follows that
\begin{align}
	\frac{1}{N^2} \sum_{i,j} \E_i \circ \E_j (\rho)&= \frac{1}{N}  \sum_i \E_i \Big(\frac{1}{N}\sum_j \E_j (\rho)\Big) \notag \\
	&= \frac{1}{N}  \sum_i \E_i \Big(\frac{1}{N}\sum_j U_j^\dagger \E (U_j \rho U_j^\dagger) U_j\Big) \notag \\
	& =  \frac{1}{N}  \sum_i U_i ^\dagger \E\Big( U_i \D_{\eta_\E} (\rho) U_i^\dagger \Big)U_i \notag \\	
	&= \D_{\eta_\E} ( \D_{\eta_\E} (\rho) )\,,
\end{align}
where we have used the linearity of channels.
What this means is that by making all possible combinations of $\E_i$ and $\E_j$ on different copies of the state and the channels, and averaging over them, we can create the composition of the two twirled channels. As for the way to compute the switch of the twirled channels in practice, this can be done by:
\begin{enumerate}
	\item producing $N^2$ copies of the state,
	\item for each value of $i,j$ combining the channels $\E_i$ and $\E_j$ in a quantum switch, 
	\item sending each copy through a switch $\Sw(\E_i, \E_j)$
	\item using each copy for whatever purpose (e.g. measuring) and
	\item averaging outcomes over all $N^2$ copies.
\end{enumerate} 
Note, however, that there is the following complication. As a measurement is made on the ancilla for each of the $N^2$ copies, same measurement outcomes should be grouped and averaged over.
That means that for each value of $i,j$ at least one $+$ outcome needs to be obtained before the averaging to produce the $C_+$ channel, and similarly for $-$. Thus, given the randomness of the outcomes after measurements in the ancilla, more repetitions are necessary. For instance to replicate the $C_+$ channel one would need to keep repeating the process for each $i,j$ until the $+$ outcome is obtained, which clearly has a high resource cost.

However, irrespective of its inefficiency, this effectively evaluates the quantum switch of the two depolarisation channels that follow from twirling, i.e. $\Sw(\D_{\eta_\E},\D_{\eta_\E})(\rho)$. Thus, results on the depolarisation channel can be effectively applied to any channel, however, at the significant resource cost of multiple copies of the state $\rho$, the channel $\E$, and the need for one switch for each copy of the state. Specifically, $N^2$ copies of the state, $2N^2$ copies of the channel, as well as $N^2$ quantum switches. For example, for a general unknown channel in $d=2$, assuming the existence of a minimal unitary 2-design with 10 elements, one would need to implement $100$ switches in parallel. In practice, however, this number can be significantly smaller if the class of channels is restricted.  That is, channel-specific twirling in some cases exists that can be used to significantly reduce this overhead. For instance, a Pauli channel can be twirled with only three unitary matrices \cite{kechrimparis_measurementprotected_2019}, significantly reducing the cost from 10. In this case, only 9 copies of the state, 18 copies of the channel and 9 quantum switches are sufficient.

The generalisation to the case of a superswitch of order $n$ is straightforward. The $n$-order superswitch combines $2^{n}$ copies of the channel and one can twirl and thus depolarise each of them by:
\begin{enumerate}
	\item producing $N^{2^{n}}$ copies of the state,
	\item for each value of the ${2^{n}}$ indices $i_1,\ldots,i_{2^{n}}$ combining the channels $\E_{i_1}, \ldots,\E_{i_{2^{n}}}$ in an $n$-order superswitch, 
	\item sending each copy through the superswitch $\Sw^{(n)}(\E_{i_1}, \ldots,\E_{i_{2^{n}}})$
	\item using each copy for whatever purpose (e.g. measuring) and
	\item averaging outcomes over all $N^{2^{n}}$ copies.
\end{enumerate} 
This effectively evaluates the $n$-order superswitch of the $2^{n}$ depolarisation channels that follow from twirling, i.e. $\Sw^{(n)}(\D_{\eta_\E},\ldots,\D_{\eta_\E})(\rho)$. Clearly, this becomes impractical very quickly as $N^{2^{n}}$ copies of the state, $2^{n}N^{2^{n}}$ copies of the channel, as well as $N^{2^{n}}$ $n$-order superswitches are necessary to implement this. However, once again this overhead may be reduced by channel-specific twirling.

Finally, in view of the main results, to distil the channel, it is only necessary to implement the second-order superswitch and obtain the sequence of outcomes `$- - +$'. Thus, for channel distillation, a minimal protocol is given in the following steps:
\begin{enumerate}
	\item Produce a copy of the state $\rho$ to be sent over the channel.
	\item For each value of the four indices $i,j,k,l$ combine the channels $\E_{i}, \E_{j},\E_{k},\E_{l}$ in a first-order superswitch and measure the three ancillas.
	\item If the obtained outcome is different than `$- - +$', repeat. Otherwise, use the obtained state for whatever purpose (e.g. measuring) and proceed to a new combination of values of $i,j,k,l$.
	\item When all combinations of $i,j,k,l$ have been exhausted and an outcome `$- - +$' has been obtained for each one of them, average the outcomes on the states over all copies with different indices, thus obtaining the original state $\rho$.
\end{enumerate} 

\section{Results 1 and 2 do not hold in $d>2$} 
\label{app: Results 1 and 2 do not hold in d>2}
In this section we show that our results do not hold for all unital channels in any dimension $d=2^n$ with $n>1$. We recall that the reason that this was possible in $d=2$ was based on two facts: (i) for any channel a minus result on the ancillas gives a channel with $p_0=0$, and (ii) for any Pauli channel with $p_0=0$, a plus outcome leads to the identity channel. We will show that while (i) still holds, (ii) fails. 

The general expression for the superswitches, Eq.\@ \eqref{eq: general switch}, involves terms with nested commutators and anticommutators of Kraus operators of the channels. A recursive procedure to evaluate higher-order superswitches from lower-order ones was given in \cite{kechrimparis_enhancing_2024}. Effectively, commutators and anti-commutators of Kraus operators of a Pauli channel map to another Pauli channel, as in the expression for the quantum switch, Eq.\@ \eqref{eq: state quantum switch commutators}.

In order to evaluate the resulting channels, let $\E(\rho)=\sum_i \vec{p}_{1,i} \sigma_i \rho \sigma_i$ and $\F(\rho)=\sum_i \vec{p}_{2,i} \sigma_i \rho \sigma_i$, denote two Pauli channels with probability vectors $\vec{p}_i = \{\alpha_i, \beta_i, \gamma_i, \delta_i \}$ where $\vec{\sigma}=\{\Id,X,Y,X\}$ is the vector of Pauli matrices. We gather all anticommutators and commutators of Kraus operators in two matrices with operator-valued elements, $\mathcal{A}[\E,\F]$ and $\mathcal{C}[\E,\F]$, respectively, given by
\begin{align}
	\mathcal{A}[\E,\F]= 2
	\begin{pmatrix}
		\sqrt{\alpha_1 \alpha_2} \, \Id & \sqrt{\alpha_1 \beta_2} \, X & \sqrt{\alpha_1 \gamma_2} \, Y & \sqrt{\alpha_1 \delta_2} \, Z \\
		\sqrt{\beta_1 \alpha_2} \, X & \sqrt{\beta_1 \beta_2} \, \Id &0 & 0 \\
		\sqrt{\gamma_1 \alpha_2} \, Y & 0 & \sqrt{\gamma_1 \gamma_2} \, \Id & 0 \\
		\sqrt{\delta_1 \alpha_2} \, Z &0 & 0 & \sqrt{\delta_1 \delta_2} \, \Id  \label{eq: acom matrix}
	\end{pmatrix} \,,
\end{align}
and
\begin{align}
	\mathcal{C}[\E,\F]= 2 \textrm{i} 
	\begin{pmatrix}
		0 & 0 & 0 & 0 \\
		0 & 0 & \,\,\,\,\,\sqrt{\beta_1 \gamma_2} \, Z & -\sqrt{\beta_1 \delta_2} \, Y \\
		0 & -\sqrt{\gamma_1 \beta_2} \, Z  & 0 & \,\,\,\,\,\,\sqrt{\gamma_1 \delta_2} \, X \\
		0 & \,\,\,\,\,\,\sqrt{\delta_1 \beta_2} \, Y & -\sqrt{ \delta_1 \gamma_2} \, X & 0 
	\end{pmatrix} \,.
\end{align}
Then, the two terms in the definition of the quantum switch, Eq.\@ \eqref{eq: zero order ss- coms and anticoms}, are equivalent to
\begin{align}
	\mathfrak{a}[\E,\F]&\equiv\frac{1}{4} \sum_{i,j} \left\{E_i, F_j \right\}\rho\left\{E_i, F_j \right\}^\dagger \notag \\
	 &=e^\top \cdot \left( \mathcal{A}[\E,\F] \circ \rho \circ \mathcal{A}[\E,\F]^* \right) \cdot e\,, \notag \\
	\mathfrak{c}[\E,\F]&\equiv\frac{1}{4} \sum_{i,j} \left[E_i, F_j\right]\rho \left[E_i, F_j\right] ^\dagger \notag \\
	 &=e^\top \cdot \left( \mathcal{C}[\E,\F] \circ \rho \circ \mathcal{C}[\E,\F]^* \right) \cdot e \,,
\end{align}
where $e=(1,1,1,1)$ denotes a vector of ones and `$\cdot$' denotes usual matrix multiplication. We have introduced two operations: (i) `$\ldots \circ \ldots \circ \ldots$' that takes two $4\times 4$ operator-valued matrices on the left and right side and a density matrix in the middle and returns a $4\times 4$ operator-valued matrix, and (ii) `$^{*}$' which denotes taking the conjugate transpose of each element of the operator-valued matrix. In detail, the first operation consists of element-wise multiplication of the two outside matrices while `sandwiching' the density matrix $\rho$ with the operators on the left and right, keeping the order. For example, the third element of the first row of $\mathcal{A} \circ \rho \circ \mathcal{A}$ would give the contribution $\left[\mathcal{A} \circ \rho \circ \mathcal{A}^*\right]_{13}= \alpha_1 \gamma _2 Y\rho Y$. The other operation, $^*$, is unimportant in the case we are considering, as conjugate transposes of Pauli matrices leave them unchanged. 
By taking left and right matrix products with the vector with unit entries $e$, we effectively get all terms in the summation. With the introduction of this concise notation, we readily find
\begin{align}
	\mathfrak{a}[\E,\F] &= \left(\alpha_1 \alpha_2 +\beta_1\beta_2+\gamma_1 \gamma_2 +\delta_1 \delta_2\right) \rho \notag \\
    &\hspace{5mm} + \left(\alpha_1 \beta_2+\beta_1 \alpha_2\right)X\rho X 
	 + \left(\alpha_1 \gamma_2+\gamma_1 \alpha_2\right)Y\rho Y\notag \\
     &\hspace{5mm}+ \left(\alpha_1 \delta_2+\delta_1 \alpha_2\right)Z\rho Z \,, \notag \\ 
		\mathfrak{c}[\E,\F]&=
	 \left( \gamma_1 \delta_2+\delta_1 \gamma_2\right)X\rho X +\left(\delta_1 \beta_2+\beta_1 \delta_2\right)Y\rho Y\notag \\
	&\hspace{5mm}+ \left( \beta_1 \gamma_2+\gamma_1 \beta_2\right)Z\rho Z \,. \label{eq: com-acom two channels}
\end{align}
Note that the expressions do not give channels but channels multiplied by probabilities.

The generalisation to higher-order superswitches is straightforward. For instance, the channel of interest corresponding to the outcomes `$--+$' in the case of the second-order superswitch is
\begin{align}
	&\frac{1}{64} \sum_{i,j,k,l} \left\{[E_i, F_j] , [\ee_k, \ff_l]  \right\}\rho\left\{[E_i, F_j] , [\ee_k, \ff_l]  \right\}^\dagger =\notag \\
	&e^\top \cdot \left( \mathcal{A}\big[\mathfrak{c}[\E,\F],\mathfrak{c}[\EE,\FF]\big] \circ \rho \circ \mathcal{A}\big[\mathfrak{c}[\E,\F],\mathfrak{c}[\EE,\FF]\big]^* \right) \cdot e \,.
\end{align}
Letting $\E=\F=\EE=\FF=\sum_i p_i \sigma_i \rho\sigma_i $, we readily obtain
\begin{align}
	\symsc{C^{(2)}}{--+}= 4(p_1^2 p_2^2+p_2^2 p_3^2+p_3^2 p_1^2) \rho = \q{--+}^{(2)}\, \id\,.
\end{align}
A crucial feature for this is the fact that for two channels with $\alpha_1 =\alpha_2=0$,  the only terms remaining in the matrix of anti-commutators $\mathcal{A}[\E,\F]$ in Eq.\@ \eqref{eq: acom matrix}, involve the identity matrix. However, when moving to Pauli channels in dimensions $2^l$ with $l>1$, all Kraus operators involve operators of the form $\sigma_i \otimes \sigma_j\otimes\ldots$. This implies that the matrix of anticommutators $	\mathfrak{a}[\E,\F]$ does not have the simple form of Eq.\@ \eqref{eq: acom matrix}. Specifically, even when the portion of the identity of the two channels $\E,\F$ is equal to zero, there will remain terms that involve operators that are not proportional to $\Id\otimes \Id \cdots\otimes \Id$. For instance, in $d=4$ we have the anticommutator 
\begin{align}
	\{\Id\otimes X, X \otimes \Id\}=2X \otimes X \label{eq: fail anticom}
\end{align}
as well as similar terms with $X$ replaced with $Y$ or $Z$, among others. Thus, it is clear that the crucial feature that enabled distillation to the identity channel in $d=2$, fails to hold in $d=4$. Moreover, such terms are also included trivially in any dimension $d=2^n$ with any $n>2$ as, for example
\begin{align}
	&\hspace{-10mm}\{\Id\otimes\ldots\otimes \Id\otimes X,\Id\otimes\ldots\otimes X \otimes \Id \} \notag \\
	&= \Id\otimes\ldots\otimes \{\Id\otimes X, X \otimes \Id \} \notag \\
    &=2\left(\Id\otimes\ldots\otimes X \otimes X\right)\,.
\end{align}
It follows that in any dimension $d\neq 2$, Results 1 and 2 do not hold for all Pauli channels.

\section{Recurrence Relations and Asymptotic Distillation Rate}
\label{app: Recurrence Relations and Asymptotic Distillation Rate}
\subsection{Channels in the interior of the tetrahedron}

In this section we show that the distillation rate after the action of a superswitch of order $n$ on $2^n$ copies of a Pauli channel $P_{\vec{p}}$, can be evaluated given knowledge of the action of the superswitch at order $n-1$. Thus, we construct recurrence relations that can be iterated to find the action of the superswitch at any order. In addition, by analysing the fixed points of the recurrence relations, we can find the asymptotic behaviour that follows from the superswitches and thus obtain the asymptotic distillation rate for certain channels.

We use the following definitions. We denote a Pauli channel with $\P_{\vec{p}} (\rho)=p_0\rho+ p_1 X\rho X+ p_2Y \rho Y+ p_3 Z\rho Z$ and for notational simplicity we may represent it by its probability vector $\vec{p}=\{p_0,p_1,p_2,p_3\}$. We denote with $\N_{\vec{q}}$ a channel that has zero portion of the identity, that is, the channel $\N_{\vec{q}} (\rho)= q_1 X\rho X+ q_2 Y \rho Y+ q_3 Z\rho Z$. We recall Eq.\@ \eqref{eq: com-acom two channels} of Appendix \ref{app: Results 1 and 2 do not hold in d>2}, and rewrite concisely in terms of the two vectors $\vec{p}$ and $\vec{q}$ of two Pauli channels $\E_{\vec{p}}$ and $\F_{\vec{q}}$ as
\begin{align}
	\mathfrak{a}[\E,\F]= \Pr(\mathfrak{a})
	\Big\{ & \frac{\sum_{i=0}^3 p_i q_i}{\Pr(\mathfrak{a})} , \frac{p_0 q_1+p_1 q_0}{\Pr(\mathfrak{a})}, \notag \\
	& \, \, \,  \frac{p_0 q_2+p_2 q_0}{\Pr(\mathfrak{a})},\frac{p_0 q_3+p_3 q_0 }{\Pr(\mathfrak{a})} \Big\} \,, \notag \\ 
		\mathfrak{c}[\E,\F]=  \Pr(\mathfrak{c})
	\Big\{ & 0 , \frac{p_1 q_2+p_2 q_1}{\Pr(\mathfrak{c})}, \frac{p_2 q_3+p_3 q_2}{\Pr(\mathfrak{c})}, \notag \\
	& \hspace{20mm} \frac{p_3 q_1+p_1 q_3 }{\Pr(\mathfrak{c})} \Big\}  \,, \label{eq: com-acom two channels vec form}
\end{align}
where
\begin{align}
    \Pr(\mathfrak{c}) &=p_1 q_2+p_2 q_1+p_2 q_3+p_3 q_2+p_3 q_1+p_1 q_3 \,, \notag \\
    \Pr(\mathfrak{a}) &= 1-  \Pr(\mathfrak{c}) \,.
\end{align}
In turn, we can rewrite this as
\begin{align}
	\mathfrak{a}[\E,\F]= \Pr(\mathfrak{a}) \P_{\vec{f}(\vec{p},\vec{q})} \,\, \,, \, 
	\mathfrak{c}[\E,\F]=  \Pr(\mathfrak{c}) \N_{\vec{g}(\vec{p},\vec{q})}
\end{align}
with
\begin{align}
    \vec{f}(\vec{p},\vec{q}) & = \frac{1}{\Pr(\mathfrak{a})} \Big\{ \sum_{i=0}^3 p_i q_i , \, p_0 q_1+p_1 q_0, \notag \\
	& \hspace{18mm}  p_0 q_2+p_2 q_0, \, p_0 q_3+p_3 q_0  \Big\} \,, \notag \\
 \vec{g}(\vec{p},\vec{q}) & = \frac{1}{\Pr(\mathfrak{c})} \Big\{  0 , \,p_1 q_2+p_2 q_1,\, p_2 q_3+p_3 q_2, \notag \\
	& \hspace{20mm} p_3 q_1+p_1 q_3  \Big\}\,. 
\end{align}
We see that when two Pauli channels are combined in a quantum switch, the anticommutator term that corresponds to the `$+$' outcome after a measurement on the ancilla leads to an updated Pauli channel with probability vector $\vec{f}(\vec{p},\vec{q})$, while the commutator term that corresponds to the `$-$' outcome after a measurement on the ancilla leads to a Pauli channel with probability vector $\vec{g}(\vec{p},\vec{q})$, which has zero portion of the identity. In addition, the identity channel can only be obtained if $p_0=q_0=0$. This means that at each order only outcomes from the previous order that lead to channels with zero portion of the identity can lead to the identity at the current order. As a result, at order $n$ we need to track and group together three terms that lead to different behaviours:
\begin{itemize}
    \item Outcomes that lead to the identity channel, $\id$.
    \item Outcomes that lead to channels with zero portion of the identity, $\N_{\vec{q}^{\,(n)}}$.
    \item Outcomes that lead to Pauli channels that do not belong in either of the two aforementioned classes, $\P_{\vec{p}^{\,(n)}}$.
\end{itemize}
The total resulting channel at order $n$ of the superswitch can be written as
\begin{align}
    \Sw^{(n)} =& \alpha^{(n)} \id \otimes \rho_{\id}+\beta^{(n)} \N_{\vec{q}^{\,(n)}}  \otimes   \rho_{\N}  \notag \\
    &+\gamma^{(n)} \P_{\vec{p}^{\,(n)}}  \otimes   \rho_{\P} \,. \label{eq: effective superswitch}
\end{align}
By $\rho_{\id}, \rho_{\N}, \rho_{\P}$ we denote the unnormalised states with orthogonal supports that gather the outcomes $o_{\id}, o_{\N}, o_{\P}$ of measurements on the ancillas that lead to the identity channel $\id$, a channel with zero portion of the identity $\N$, i.e. $q^{(n)}_{0}=0$, and a general Pauli channel with $p^{(n)}_{0} \neq 0,1$, respectively. For instance, in the case of the first-order superswitch, it follows from Eq.\@ \eqref{eq: channels 1st} that 
\begin{align}
    \rho_{\id} &= \ketbra{--+}\,,  \notag \\
    \rho_{\N}&=\ketbra{++-}+\ketbra{-+-}+\notag \\
    &+\ketbra{+--}+\ketbra{---}\,,\notag \\
    \rho_{\P}&=\Id -\rho_{\id}-\rho_{\N} = \big( \ketbra{+++} \notag \\
    &+\ketbra{-++}+\ketbra{+-+} \,.
\end{align}
The fact that it is possible to summarise all different channels into essentially three channels, follows from the fact that for two Pauli channels $\P_{\vec{p}}$ and $\P_{\vec{q}}$
\begin{align}
    \sum_i \lambda_i \P_{\vec{p}_i} = \left(\sum_i \lambda_i\right) \P_{\vec{p}^\prime} \, \,, \,\, \vec{p}^{\prime}=\frac{\sum_i \lambda_i \vec{p}_i}{\sum_i \lambda_i}\,.
\end{align}

Given that superswitches are switches of switches and that the action of the superswitch at order $n$ is assumed to take the form in Eq.\@ \eqref{eq: effective superswitch} we can explicitly evaluate the superswitch at order $n+1$. In practice, that means that for each combination of terms we need to evaluate the contribution from the anticommutator and commutator contribution from Eq.\@ \eqref{eq: iteration superswitches}. There are nine possible combinations, each of which leads to two possible outcomes so in principle we will have 18 terms at order $n+1$. However, by noticing that combining the identity with any channel leads only to one outcome, i.e. the channel itself, and that the ordering of combining two channels, i.e. $\mathfrak{a}[\E,\F]=\mathfrak{a}[\F,\E]$ and $\mathfrak{c}[\E,\F]=\mathfrak{c}[\F,\E]$ does not matter, we find nine unique terms. Explicitly,
\begin{align}
    &\Sw^{(n+1)} = \left(\left(\alpha^{(n)}\right)^2 +\frac{\left(\beta^{(n)}\right)^2}{3} \right)\id \otimes \rho_{\id}\notag \\ 
    &+ \bigg( 2\alpha^{(n)}\beta^{(n)}  \N_{\vec{q}^{\,(n)}} + \frac{2\left(\beta^{(n)}\right)^2}{3} \N_{\vec{q}_{\nicefrac{4}{3}}}+ \notag \\
    &\hspace{6mm}+\left(\gamma^{(n)}\right)^2 \Pr(\mathfrak{c}[\vec{p}^{\,(n)},\vec{p}^{\,(n)}]) \N_{\vec{g}(\vec{p}^{\,(n)},\vec{p}^{\,(n)})} \notag \\
    &\hspace{6mm}+2\beta^{(n)} \gamma^{(n)} \Pr(\mathfrak{c}[\vec{p}^{\,(n)},\vec{q}^{\,(n)}]) \N_{\vec{g}(\vec{p}^{\,(n)},\vec{q}^{\,(n)})}\bigg) \otimes   \rho_{\N}  \notag \\
    &+\bigg(2 \beta^{(n)}\gamma^{(n)} \Pr(\mathfrak{a}[\vec{p}^{\,(n)},\vec{q}^{\,(n)}]) \P_{\vec{f}(\vec{p}^{\,(n)},\vec{q}^{\,(n)})}  \notag \\
    &\hspace{6mm}+\left(\gamma^{(n)}\right)^2 \Pr(\mathfrak{a}[\vec{p}^{\,(n)},\vec{p}^{\,(n)}]) \P_{\vec{f}(\vec{p}^{(n)},\vec{p}^{(n)})} \notag \\
    &\hspace{6mm}+2\alpha^{(n)} \gamma^{(n)} \P_{\vec{p}^{\,(n)}} \bigg) \otimes   \rho_{\P}\,,  \label{eq: superswitch n+1}
\end{align}
where $\vec{q}_{\nicefrac{4}{3}}=\{0,\nicefrac{1}{3},\nicefrac{1}{3},\nicefrac{1}{3}\}$.
The last expression is of the form
\begin{align}
    \Sw^{(n+1)} =& \alpha^{(n+1)} \id \otimes \rho_{\id}+\beta^{(n+1)} \N_{\vec{q}^{\,(n+1)}}  \otimes   \rho_{\N}  \notag \\
    &+\gamma^{(n+1)} \P_{\vec{p}^{\,(n+1)}}  \otimes   \rho_{\P}
    \,. \label{eq: effective superswitch n+1}
\end{align}
where
\begin{align}
    \alpha^{(n+1)} &= \left(\alpha^{(n)}\right)^2 +\frac{\left(\beta^{(n)}\right)^2}{3} \,, \notag \\
    \beta^{(n+1)} &= 2\alpha^{(n)}\beta^{(n)} + \frac{2\left(\beta^{(n)}\right)^2}{3} \notag \\
    &\hspace{4mm}+ \left(\gamma^{(n)}\right)^2 \Pr(\mathfrak{c}[\vec{p}^{\,(n)},\vec{p}^{\,(n)}]) \notag \\
    &\hspace{4mm}+2\beta^{(n)} \gamma^{(n)} \Pr(\mathfrak{c}[\vec{p}^{\,(n)},\vec{q}^{\,(n)}]) \,, \notag \\
    \gamma^{(n+1)} &= 2\alpha^{(n)} \gamma^{(n)} +2 \beta^{(n)}\gamma^{(n)} \Pr(\mathfrak{a}[\vec{p}^{\,(n)},\vec{q}^{\,(n)}])  \notag \\
    &\hspace{4mm}+\left(\gamma^{(n)}\right)^2 \Pr(\mathfrak{a}[\vec{p}^{\,(n)},\vec{p}^{\,(n)}]) \,, \notag \\
    \vec{q}^{\,(n+1)} &= \bigg(2\beta^{(n)} \gamma^{(n)} \Pr(\mathfrak{c}[\vec{p}^{\,(n)},\vec{q}^{\,(n)}]) \vec{g}(\vec{p}^{\,(n)},\vec{q}^{\,(n)}) \notag \\
    &\hspace{6mm}+\left(\gamma^{(n)}\right)^2 \Pr(\mathfrak{c}[\vec{p}^{\,(n)},\vec{p}^{\,(n)}]) \vec{g}(\vec{p}^{\,(n)},\vec{p}^{\,(n)})\notag \\
    &\hspace{6mm}+ 2\alpha^{(n)}\beta^{(n)}  \vec{q}^{\,(n)} + \frac{2\left(\beta^{(n)}\right)^2}{3} \vec{q}_{\nicefrac{4}{3}}\bigg)\big/\beta^{(n+1)} \,, \notag \\
    \vec{p}^{\,(n+1)}&= \bigg(2 \beta^{(n)}\gamma^{(n)} \Pr(\mathfrak{a}[\vec{p}^{\,(n)},\vec{q}^{\,(n)}]) \vec{f}(\vec{p}^{\,(n)},\vec{q}^{\,(n)}) \notag \\
    &\hspace{6mm}+\left(\gamma^{(n)}\right)^2 \Pr(\mathfrak{a}[\vec{p}^{\,(n)},\vec{p}^{\,(n)}]) \vec{f}(\vec{p}^{\,(n)},\vec{p}^{\,(n)}) \notag \\
    &\hspace{6mm} +2\alpha^{(n)} \gamma^{(n)} \vec{p}^{\,(n)} \bigg) \big/ \gamma^{(n+1)} \,. \label{eq: recurrence relations general}
\end{align}
These equations effectively define recurrence relations that can be iterated to find the action of a superswitch of any order. The initial conditions for the recurrence relations are $\alpha^{(0)}=\beta^{(0)}=0,\gamma^{(0)}=1,\vec{q}^{\,(0)}=0,\vec{p}^{\,(0)}=\vec{p}$. Note that since the zeroth-order superswitch is the channel $\P_{\vec{p}}$ only without any superswitch acting, the index $n$ starts from the value $0$. We highlight that the parameter $\alpha^{(n)}$ gives the distillation rate at order $n$, which is zero in general for $n=1$ (aside from channels with $p_0=0$). Evaluating the second-order superswitch by iterating we find $\R^{(2)}=\alpha^{(2)} = 4 (p_1 p_2 + p_2 p_3 + p_3 p_1)$. Higher orders can be evaluated iteratively, but we do not list the expressions as they become cumbersome.

Instead, we focus on a particular example, the depolarisation channel. In this case, it can be seen that the parameters $\vec{q}^{\,(n)}$ remain constant for any value of $n$ and take the value $\vec{q}^{\,(n)}=\vec{q}_{\nicefrac{4}{3}}=\{0,\nicefrac{1}{3},\nicefrac{1}{3},\nicefrac{1}{3}\}$. In addition, there is only one recurrence relation for $\vec{p}^{\,(n)}$ as the Pauli vector is that of a depolarisation channel. Thus, the recurrence relations are significantly simplified. The distillation rates, $\R^{(n)}=\alpha^{(n)}$, can be obtained in a straightforward manner and the expressions for orders of up to five are given in Eqs.\@ \eqref{eq: distil rate 1} and \eqref{eq: distil rate 2}. Orders of up to eight are plotted in Fig.\@ \ref{fig: communication rates}.

Interestingly, in this case we can also analyse the asymptotic behaviour by looking for the fixed points of the recurrence relations. These follow by setting
\begin{align}
    \alpha^{(n+1)}=\alpha^{(n)} &:= \alpha\,, \, \beta^{(n+1)}=\beta^{(n)}:=\beta \,, \notag \\
    \gamma^{(n+1)}= \gamma^{(n)} &:= \gamma \,, \notag \\
    \vec{p}^{\,(n+1)}= \vec{\,p}^{(n)} &:= \left\{1-\frac{3p}{4},\frac{p}{4},\frac{p}{4},\frac{p}{4}\right\}\,,
\end{align}
and explicitly we obtain
\begin{align}
    \alpha &= \alpha^2 +\frac{\beta^2}{3} \,, \notag \\
    \beta &= 2\alpha\beta +\frac{2\beta^2}{3}+4\beta\gamma p +6\gamma^2 p^2 \,,\notag \\
    \gamma&=2\alpha\gamma +2\beta\gamma (1-2p)+\gamma^2 (1-6p^2)  \,, \notag \\
    p&=\left(2\alpha\gamma p +2\beta\gamma \left(\frac{1}{3}-p\right)+2\gamma^2 (1-3p)p \right) /\gamma \,.
\end{align}
All solutions to these equations can be analytically obtained. Some correspond to the trivial solution of starting with the identity channel, $p=0$. Another solution corresponds to the case of starting with the depolarisation channel with $p=\nicefrac{4}{3}$,
\begin{align}
    \alpha_{\nicefrac{4}{3}}^{(\infty)} = \frac{1}{4}\,, \, \beta_{\nicefrac{4}{3}}^{(\infty)} &= \frac{3}{4} \,, \, \gamma_{\nicefrac{4}{3}}^{(\infty)}=0 \,, \notag \\
     p_{\nicefrac{4}{3}}^{(\infty)} &= \frac{3-\sqrt{3}}{6} \,.
\end{align}
The value of $\alpha^{(\infty)}_{\nicefrac{4}{3}}$ gives the asymptotic distillation rate, $\R^{(\infty)}_{\nicefrac{4}{3}} = \alpha_{\nicefrac{4}{3}}^{(\infty)} =\nicefrac{1}{4}$. We note that this is the only case of a depolarisation channel (excluding the trivial identity channel) where the quantum switch achieves a nonzero distillation rate, which is found to be $\R^{(1)}_{\nicefrac{4}{3}}=\nicefrac{1}{3}$. With each subsequent superswitch the distillation rate is decreasing, tending to the value $\R^{(\infty)}_{\nicefrac{4}{3}}$ from above.

The general case corresponding to starting with a depolarisation channel with any value $p\in ( 0,\nicefrac{4}{3})$, is given by the solution
\begin{align}
    \alpha^{(\infty)} = \frac{2-\sqrt{3}}{4}\,, \, \beta^{(\infty)} &= \frac{\sqrt{3}}{4} \,, \, \gamma^{(\infty)}=\frac{1}{2} \,, \notag \\ p^{(\infty)} &= \frac{3-\sqrt{3}}{6} \,. \label{eq: solution fixed points}
\end{align}
Thus the asymptotic distillation rate for any depolarisation channel with $p\neq 0,\nicefrac{4}{3}$ is $\R^{(\infty)}_{\text{dep}}=\nicefrac{(2-\sqrt{3})}{4}\approx 0.067$.
The fixed point solution tends to
\begin{align}
    \Sw^{(\infty)} =& \left(\frac{2-\sqrt{3}}{4}\right) \id \otimes \rho_{\id}+\left(\frac{\sqrt{3}}{4}\right) \D_{\nicefrac{4}{3}}  \otimes   \rho_{\N}  \notag \\
    &+\frac{1}{2} \D_{\frac{3-\sqrt{3}}{6}}  \otimes   \rho_{\P}
    \,.
\end{align}
Interestingly, tracing out the ancilla we find that the fixed point corresponds to a completely depolarising channel since in terms of the probability vector $\vec{p}$ we have
\begin{align}
    \Sw^{(\infty)} =& \left(\frac{2-\sqrt{3}}{4}\right) \{1,0,0,0\}+\left(\frac{\sqrt{3}}{4}\right) \{0,\nicefrac{1}{3},\nicefrac{1}{3},\nicefrac{1}{3}\}  \notag \\
    &+\frac{1}{2} \left\{1-\frac{3p}{4},\frac{p}{4},\frac{p}{4},\frac{p}{4}\right\} = \left\{\frac{1}{4},\frac{1}{4},\frac{1}{4},\frac{1}{4}\right\} \,.
\end{align}

The fixed points for the general case, Eqs.\@ \eqref{eq: recurrence relations general}, are identical to the case of the depolarisation channel. Excluding points on the faces of the tetrahedron, we find that the only solution is $\vec{p}^{\,(\infty)} = \{1-\nicefrac{3p^{(\infty)}}{4},\nicefrac{p^{(\infty)}}{4},\nicefrac{p^{(\infty)}}{4},\nicefrac{p^{(\infty)}}{4}\}$ and $\vec{q}=\{0,\nicefrac{1}{3},\nicefrac{1}{3},\nicefrac{1}{3}\}$ and the rest of the parameters identical to the one given by Eqs.\@ \eqref{eq: solution fixed points}. It turns out that in the general case, after a few iterations, the initial arbitrary vector $\vec{q} ^{\,(n)}$ quickly tends to the vector $\{0,\nicefrac{1}{3},\nicefrac{1}{3},\nicefrac{1}{3}\}$, while the vector $\vec{p}^{\,(n)}$ tends to a probability vector of a depolarisation channel and thus the general case quickly converges to the case of the depolarisation channels. Thus, the asymptotic distillation rate for any channel inside the tetrahedron is $\R^{(\infty)}_{\text{in}}=\nicefrac{(2-\sqrt{3})}{4}\approx 0.067$

\subsection{Channels on the faces of the tetrahedron}
From the four faces of the tetrahedron, we split the discussion into two cases, since the recurrence relations slightly differ:
\begin{itemize}
    \item The three faces consisting of Pauli channels $\P_{\vec{p}}$ with $p_0\neq 0$ and exactly one $p_i=0\,, i\in\{1,2,3\}$.
    \item The face consisting of Pauli channels $\P_{\vec{p}}$ with $p_0=0$ and $p_i\neq 0 \,, \forall i\in\{1,2,3\}$, that is, the base of the tetrahedron.
\end{itemize}

\subsubsection{The three faces consisting of Pauli channels $\P_{\vec{p}}$ with $p_0\neq 0$ and exactly one $p_i=0\,, i\in\{1,2,3\}$.}

Without loss of generality, we will consider the case of the face consisting of channels of the form $\P_{\vec{p}}=(1-p-q)\rho + p X \rho X +q Y \rho Y$. The derivation for the other two cases is identical.

In contrast to the case of channels inside the tetrahedron, channels on the faces can also lead to unitary channels after combining in the quantum switch. Thus, unitary contributions must also be tracked at each order.
As a result, at order $n$ we need to track the four terms that lead to different behaviours: the three terms that were tracked for channels inside the tetrahedron, as well as an extra term that tracks a unitary channel. Specifically, we have
\begin{itemize}
    \item Outcomes that lead to the identity channel, $\id$.
    \item Outcomes that lead to the unitary channel $\U_Z = Z(\cdot)Z$.
    \item Outcomes that lead to channels with zero portion of the identity, $\M_{\vec{q}^{\,(n)}}$, with $\vec{q}^{\,(n)}\equiv\left\{0,q_1^{(n)},q_2^{(n)},0\right\}$.
    \item Outcomes that lead to Pauli channels that do not belong in either of the three aforementioned classes, $\P_{\su{\vec{p}}{\,(n)}}$, with $\vec{p}^{\,(n)}\equiv\left\{p_0^{(n)},p_1^{(n)},p_2^{(n)},0\right\}$.
    
\end{itemize}
The total channel resulting at order $n$ of the superswitches can be written as
\begin{align}
    \Sw^{(n)} \,\,=\,&\,\,\,\,\,\,  \alpha^{(n)} \id \otimes \rho_{\id}+\beta^{(n)} \U_Z \otimes \rho_{\U}  \notag \\
    &+\gamma^{(n)}\P_{\su{\vec{p}}{\,(n)}}  \otimes   \rho_{\P}  +\delta^{(n)} \M_{\su{\vec{q}}{\,(n)}}  \otimes   \rho_{\M} \,. \label{eq: effective superswitch face n}
\end{align}
By $\rho_{\id}, \rho_{\U}, \rho_{\P}, \rho_{\M}$ we denote the unnormalised states with orthogonal supports that gather the outcomes $o_{\id}, o_{\U}, o_{\P}, o_{\M}$ of measurements on the ancillas that lead to the identity channel $\id$, the unitary channel $\U_Z$, a general Pauli channel on the face $\P_{\vec{p}^{\,(n)}}$, with $\vec{p}^{\,(n)}\equiv\left\{p_0^{(n)},p_1^{(n)},p_2^{(n)},0\right\}$, and a channel with zero portion of the identity  $\M_{\vec{q}^{\,(n)}}$, with $\vec{q}^{\,(n)}\equiv\left\{0,q_1^{(n)},q_2^{(n)},0\right\}$, respectively. 

As in the case inside the tetrahedron, the next order $n+1$ follows by forming all possible combinations of channels of order $n$ from Eq.\@ \eqref{eq: iteration superswitches}. Explicitly, we find
\begin{align}
    \Sw^{(n+1)} &=\alpha^{(n+1)} \id \otimes \rho_{\id}+\beta^{(n)} \U_Z \otimes \rho_{\U}  \notag \\
    &+\gamma^{(n+1)}\P_{\su{\vec{p}}{\,(n+1)}}  \otimes   \rho_{\P}  \notag \\
    &+\delta^{(n+1)} \M_{\su{\vec{q}}{\,(n+1)}}  \otimes   \rho_{\M} \,, \label{eq: effective superswitch face n+1}
\end{align}
where
\begin{align}
    \alpha^{(n+1)} &= \left(\alpha^{(n)}\right)^2 +\left(\beta^{(n)}\right)^2 +\left(\delta^{(n)}\right)^2\left(1-2q_1^{(n)}q_2^{(n)}\right) \,, \notag \\
    \beta^{(n+1)} &= 2\left(\gamma^{(n)}\right)^2 p_1^{(n)} p_2 ^{(n)}+2\left(\delta^{(n)}\right)^2 q_1^{(n)} q_2 ^{(n)} \notag \\
    &+2\gamma^{(n)} \delta^{(n)} \left(q_1^{(n)} p_2 ^{(n)} +p_1 ^{(n)} q_2^{(n)}\right) +2\alpha^{(n)} \beta^{(n)} \,, \notag \\
    \gamma^{(n+1)} &= 2\alpha^{(n)} \gamma^{(n)} +\left(\gamma^{(n)}\right)^2 \left(1-2p_1^{(n)} p_2 ^{(n)}\right)  \notag \\
    &\hspace{4mm}+2 \gamma^{(n)} \delta ^{(n)} \left(1-q_1^{(n)} p_2 ^{(n)} -p_1 ^{(n)} q_2 ^{(n)}\right) \,, \notag \\
    \delta^{(n+1)}&=2 \alpha^{(n)} \delta^{(n)} +2\beta^{(n)} \delta^{(n)} +2 \beta^{(n)} \gamma^{(n)}\left(1-p_0^{(n)}\right)\,, \notag \\
    \vec{p}^{\,(n+1)}&= \bigg(2\gamma^{(n)}\delta^{(n)} \Pr(\mathfrak{a}[\vec{p}^{\,(n)},\vec{q}^{\,(n)}]) \vec{f}(\vec{p}^{\,(n)},\vec{q}^{\,(n)}) \notag \\
    &+\left(\gamma^{(n)}\right)^2 \Pr(\mathfrak{a}[\vec{p}^{\,(n)},\vec{p}^{\,(n)}]) \vec{f}(\vec{p}^{\,(n)},\vec{p}^{\,(n)}) \notag \\
    &+2 \alpha^{(n)} \gamma^{(n)} \vec{p} ^{\,(n)} \bigg) \big/ \gamma^{(n+1)}  \,, \notag \\
    \vec{q}^{\,(n+1)} &= \bigg(2\alpha^{(n)} \delta^{(n)} \vec{q}^{\,(n)}+2\beta^{(n)} \gamma^{(n)} \left\{0, p_2^{\,(n)},p_1^{\,(n)},0\right\}\notag \\
    &+2\beta^{(n)} \delta^{(n)} \left\{0, q_2^{\,(n)},q_1^{\,(n)},0\right\}\bigg)\big/\delta^{(n+1)} \,.\label{eq: recurrence relations general face}
\end{align}
These are the recurrence relations that can be iterated to find the action of a superswitch of any order for channels at the face of the tetrahedron. The initial conditions for the recurrence relations are $\alpha^{(0)}=\beta^{(0)}=0,\gamma^{(0)}=1,\delta^{(0)}=0,\vec{q}^{\,(0)}=0,\vec{p}^{\,(0)}=\vec{p}$. We note that in the case we are examining, the distillation rate at order $n$ is given as the sum $\alpha^{(n)}+\beta^{(n)}$.

Looking for the fixed points in the interior of the edge, we find the only solution
\begin{align}
    \alpha^{(\infty)}&=\frac{2-\sqrt{2}}{4}\,,\,\beta^{(\infty)}=\gamma^{(\infty)}=\frac{1}{4}\,,\, \delta^{(\infty)}=\frac{1}{2\sqrt{2}}\,, \notag \\
      \vec{p}^{\,(\infty)}&=\left\{\sqrt{2}-1,1-\frac{1}{\sqrt{2}},1-\frac{1}{\sqrt{2}},0\right\}\,, \notag \\
     \vec{q}^{\,(\infty)}&=\{0,\nicefrac{1}{2},\nicefrac{1}{2},0\}\,,
\end{align}
to which all points in the interior of the tetrahedron are attracted to. The asymptotic distillation rate is thus
\begin{align}
    \R^{(\infty)}_{\,\lowerrighttriangle{}}&=\alpha^{(\infty)}+\beta^{(\infty)} =\frac{2-\sqrt{2}}{4}+\frac{1}{4} \approx 0.3965\,.
\end{align}

\subsubsection{The base of the tetrahedron, consisting of Pauli channels $\P_{\vec{p}}$ with $p_0=0$ and $p_i\neq 0 \,, \forall i\in\{1,2,3\}$}
The derivation of the last face, corresponding to channels of the form $\P_{\vec{p}}= p X \rho X +q Y \rho Y +(1-p-q)Z\rho Z$ follows the derivation of the other three faces with one difference: there is no possibility for unitary channel contributions. That is, the term $\beta^{(n)} \U_Z \otimes \rho_{\U}$ does not appear in Eq.\@ \eqref{eq: effective superswitch face n}. Moreover, the general channel is a channel with zero portion of the identity and thus there is no $\M$ term, the superswitch at order $n$ will have the form
\begin{align}
    \Sw^{(n)} \,\,=\,&\,\,\,\,\,\,  \alpha^{(n)} \id \otimes \rho_{\id}+\beta^{(n)} \P_{\vec{p}^{\,(n)}}  \otimes   \rho_{\P} \,. \label{eq: effective superswitch other face n}
\end{align}
At the next order we will have
\begin{align}
    \Sw^{(n+1)} \,\,=\,&\,\,\,\,\,\,  \alpha^{(n+1)} \id \otimes \rho_{\id}+\beta^{(n+1)} \P_{\vec{p}^{\,(n+1)}}  \otimes   \rho_{\P} \,. \label{eq: effective superswitch other face n+1}
\end{align}
where
\begin{align}
    \alpha^{(n+1)} &= \left(\alpha^{(n)}\right)^2 +\left(\beta^{(n)}\right)^2\Big(1-2q_1^{(n)}q_2^{(n)}\notag \\
    &\hspace{20mm}-2q_2^{(n)}q_3^{(n)}-2q_3^{(n)}q_1^{(n)}\Big) \,, \notag \\
    \beta^{(n+1)} &= 2 \alpha^{(n)} \beta^{(n)} +\left(\beta^{(n)}\right)^2 \Big(2q_1^{(n)}q_2^{(n)}\notag \\
    &\hspace{20mm}+2q_2^{(n)}q_3^{(n)}+2q_3^{(n)}q_1^{(n)}\Big)  \,, \notag \\
    \vec{p}^{\,(n+1)}&= \frac{1}{\beta^{(n+1)}} \bigg(2\alpha^{(n)}\beta^{(n)} \vec{p}^{\,(n)} \notag \\
    &+2\left(\beta^{(n)}\right)^2\Pr(\mathfrak{c}[\vec{p}^{\,(n)},\vec{p}^{\,(n)}]) \vec{g}(\vec{p}^{\,(n)},\vec{p}^{\,(n)})  \bigg)    \,.\label{eq: recurrence relations general other face}
\end{align}
The initial conditions in this case are $\alpha^{(0)}=0,\beta^{(0)}=1,\vec{p}^{\,(0)}=\vec{p}$. Looking for the fixed points of the recurrence relations, Eq.\@ \eqref{eq: recurrence relations general other face}, we find the solution
\begin{align}
    \alpha^{(\infty)}=\frac{1}{4},\beta^{(\infty)}=\frac{3}{4},\vec{p}^{\,(\infty)}=\{0,\nicefrac{1}{3},\nicefrac{1}{3},\nicefrac{1}{3}\},
\end{align}
where we have restricted to points that are not in the edges.
Thus, the distillation rate at the base of the tetrahedron is $\R^{(\infty)}_{\text{base}}=\nicefrac{1}{4}$.

\section{Asymptotic distillation rates vs finite-order distillation rates}
\label{app: Asymptotic distillation rates vs finite-order distillation rates}
The asymptotic distillation rates derived in Appendix \ref{app: Recurrence Relations and Asymptotic Distillation Rate}, give the limiting values that can be achieved with \sw{n} as the order $n$ tends to infinity. For some channels, this implies that the higher the order of the superswitch, the higher the distillation rate. However, this is not true for all channels. In fact, there exist channels for which the highest distillation rate is achieved by the quantum switch, with each subsequent higher-order switch having a lower distillation rate, tending to the asymptotic rate from above. Channels that exhibit behaviour in between the aforementioned two cases exist as well. This mirrors a similar phenomenon emerging in the problem of quantum state discrimination \cite{kechrimparis_enhancing_2024}. 

\begin{figure}[!t]
	\includegraphics[width=1\columnwidth,trim={0.5cm 0.7cm 0 0},clip]{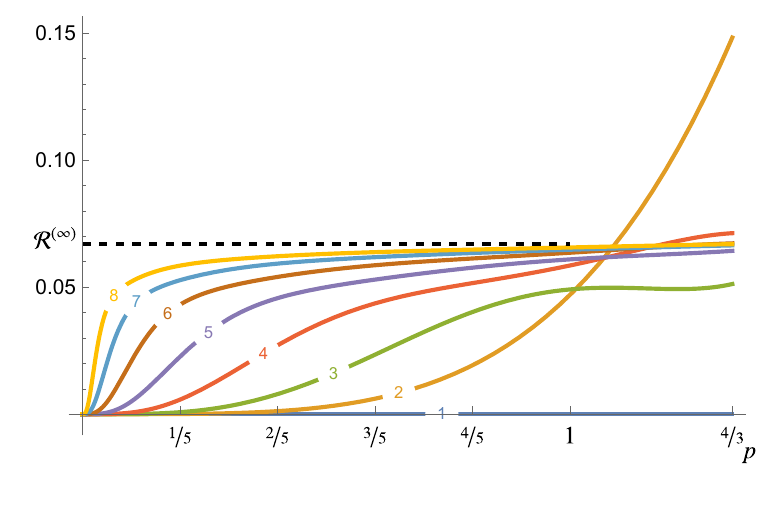}
	\caption{The distillation rates $\R^{(n)}$ achieved by the superswitches from orders one to eight. The quantum switch can not distill for any value of $p$, while each superswitch achieves a higher distillation rate for all values of $p\in[0,\nicefrac{4}{3}]$, in comparison to all previous-order superswitches.} 
    \label{fig: distillation rates dep full}
\end{figure}

As a demonstrating example, we revisit the case of the depolarisation channel $\D(\rho) = \left(1-\frac{3p}{4}\right)\rho + \frac{p}{4}\left(X\rho X + Y\rho Y + Z\rho Z \right)$ with $p\in[0,\nicefrac{4}{3}]$. In Fig.\@ \ref{fig: distillation rates dep full} we plot distillation rates with \sw{n}, $n\leq 8$. 

We note that, for the depolarisation channel, the point where the behaviour changes coincides with the point where the Bloch vectors of the input states flip direction. That is, whenever $p>1$ the Bloch vector $\vec{r}=(r_1 ,r_2, r_3)$ of a state $\rho$ is mapped to
\begin{align}
	\vec{r} = (r_1 ,r_2, r_3) \rightarrow (1-p) (r_1,r_2, r_3),
\end{align}
and thus flips direction. 

For general Pauli channels, although we do not have an exact characterisation of the different regions, we have noticed by examining a number of examples that, for channels close to the identity channel: (i) the asymptotic rate  bounds all finite orders from above, and (ii) a superswitch of order $n$ will have a higher distillation rate than all superswitches of order $m$, with $m<n$.

\end{document}